\documentclass[a4paper,11pt]{article}
\usepackage{graphicx}
\usepackage{amsmath}
\usepackage{mathtools}
\usepackage{epsf}
\usepackage{epsfig}
\usepackage{amssymb}
\usepackage{lmodern}
\usepackage{latexsym}
\usepackage{subfigure}
\usepackage{subfigure}
\usepackage{multirow}
\usepackage{txfonts}

\begin{document}
\begin{titlepage}
\vspace{.1cm}
\begin{flushright}
MZ-TH/12-17
\end{flushright}
\begin{center}

\vspace{.5cm}
{\bf\Large The Fermion Mass Hierarchy in Models with Warped Extra Dimensions and a Bulk Higgs.}\\[.3cm]
\vspace{1cm}
Paul R.~Archer\footnote{archer@uni-mainz.de}\\
\vspace{1cm} {\em  
Institute for Physics (WA THEP), Johannes Gutenberg-Universit\"{a}t,\\
D-55099 Mainz, Germany}\\[.2cm]

\end{center}
\bigskip\noindent

\begin{abstract}
The phenomenological implications of allowing the Higgs to propagate in both AdS${}_5$ and a class of asymptotically AdS spaces are considered. Without tuning, the vacuum expectation value (VEV) of the Higgs is peaked towards the IR tip of the space and hence such a scenario still offers a potential resolution to the gauge-hierarchy problem. When the exponent of the Higgs VEV is approximately two and one assumes order one Yukawa couplings, then the fermion Dirac mass term is found to range from $\sim 10^{-5}$ eV to $\sim 200$ GeV in approximate agreement with the observed fermion masses. However, this result is sensitive to the exponent of the Higgs VEV, which is a free parameter. This paper offers a number of phenomenological and theoretical motivations for considering an exponent of two to be the optimal value. In particular, the exponent is bounded from below by the Breitenlohner-Freedman bound and the requirement that the dual theory resolves the gauge hierarchy problem. While, in the model considered, if the exponent is too large, electroweak symmetry may not be broken. In addition, the holographic method is used to demonstrate, in generality, that the flatter the Higgs VEV, the smaller the contribution to the electroweak $T$ parameter. In addition, the constraints from a large class of gauge mediated and scalar mediated flavour changing neutral currents, will be at minimal values for flatter Higgs VEVs. Some initial steps are taken to investigate the physical scalar degrees of freedom that arise from a mixing between the $W_5/Z_5$ components and the Higgs components.     

\end{abstract}

\end{titlepage}
%%%%%%%%%%%%%%%%%%%%%%%%%%%%%%%%%%%%%%%%%%%%%%%%%%%%
\section{Introduction.}

In high energy physics today there exists two striking hierarchies. Firstly there exists a hierarchy between the two known dimensionful parameters, the Planck scale ($\sim10^{18}$ GeV) and the electroweak (EW) scale ($\sim 200$ GeV). Secondly there exists a hierarchy in the observed fermion masses ranging from the top mass ($\sim170$ GeV) to the lightest neutrino mass ($\sim10^{-4}-10^{-2}$ eV). Although there is considerable uncertainty, with regard to the lightest neutrino mass, it is curious that these two apparently unrelated hierarchies should extend over a similar number of orders of magnitude. It has been known for some time that warped extra dimensions, in the context of the Randall and Sundrum model (RS) \cite{Randall:1999ee}, offer a compelling explanation of such hierarchies. In particular, the fundamental scale of the model is taken to be the Planck scale and a small 4D effective EW scale is generated, via gravitational redshifting, by localising the Higgs in the IR tip of the space. Further still, by allowing the standard model (SM) particles to propagate in the bulk, fermions with zero modes peaked towards the UV tip of the space, will gain exponentially small masses \cite{Grossman:1999ra, Gherghetta:2000qt, Huber:2000ie}. While this description of flavour naturally gives rise to a large hierarchy in the fermion masses, it offers no indication of the size of that hierarchy.

As first pointed out in \cite{Agashe:2008fe}, this changes when one allows the Higgs to propagate in the bulk. The model still offers a potential resolution to the gauge hierarchy problem since it is found that in AdS${}_5$, without tuning, the Higgs profile is not flat, but peaked towards the IR tip of the space and would gain a vacuum expectation value (VEV), $h(r)\sim e^{2kr+\alpha kr}$, where $k$ is the curvature scale of the space. Assuming order one Yukawa couplings, this spatial dependence of the Higgs VEV gives rise to a `maximum' and a `minimum' fermion zero mode mass corresponding to IR and UV localised fermions. So the SM fermion masses would now extend down, from the EW scale, by a factor of $\Omega^{-1-\alpha}$. Where $\Omega \sim 10^{15}$ is the warp factor related to the difference between the EW scale and the Planck scale. For example, if $\alpha\approx 0$, then the fermion Dirac masses would range from $\sim 200$ GeV to $\sim10^{-5}$ eV. However the exponent of the Higgs VEV, $\alpha $, is essentially a free parameter. Here we argue that if it is possible to find a plausible reason for why $\alpha $ should be close to zero, then models with a bulk Higgs would help to offer some insight as to why the observed fermion mass hierarchy extends over the range that it does.  

From a purely 5D perspective of the RS model, there is no strong argument for why the Higgs should have a special status as the only brane localised particle. Since, as far as the author is aware, there are no symmetries forbidding such bulk Higgs terms. On the other hand, in a conjectured dual theory, an IR brane localised Higgs would be dual to a purely composite Higgs, whereas a bulk Higgs would be dual to a partially composite Higgs (i.e an elementary Higgs mixing with a composite sector). The corresponding Higgs operator would have a scaling dimensions of $2+\alpha$ \cite{Luty:2004ye, ArkaniHamed:2000ds, Rattazzi:2000hs, PerezVictoria:2001pa, Batell:2007ez, Gherghetta:2010cj}. Hence models with a bulk Higgs offer a convenient framework that allow for the scaling dimension of the Higgs operator to be easily varied.    

So the bulk of this paper will be concerned with investigating the phenomenological implications of changing the exponent of the Higgs VEV, $\alpha$. However there are a couple of related secondary questions. Firstly, by allowing the Higgs (a complex doublet under SU$(2)$) to propagate in the bulk, one gains additional scalar degrees of freedom that arise as a mixture of the fifth component of the $W$ and $Z$ fields and the additional components of the Higgs. Such scalar fields offer an important prediction that can help with the verification or exclusion of 5D models with a bulk Higgs. Hence in section \ref{sect:PseudoScal} we take some initial steps towards investigating the implications of such pseudo-scalars. 

A second question is related to the geometry of the space. There has been a considerable amount of work done on the phenomenology of the RS model, however this work has primarily focused on AdS${}_5$. At the very least, the RS model should include a bulk Goldberger-Wise scalar \cite{Goldberger:1999uk} and typically the model includes many more bulk fields. These bulk fields will lead to a modification in the AdS${}_5$ geometry which is assumed to be small. An important question to ask is, to what extent is the existing phenomenological analysis robust against small modifications to the geometry? With this in mind, in addition to a pure AdS${}_5$ geometry, we shall also consider a class of asymptotically AdS${}_5$ spaces that have arisen from a scalar plus gravity system \cite{Cabrer:2009we}. These spaces are of additional interest since they can, for certain regions of parameter space, result in a significant reduction in the constraints from EW observables \cite{Cabrer:2011vu, Cabrer:2011fb, Carmona:2011ib} and flavour physics \cite{Cabrer:2011qb}.     

Aside from these two additional questions, the central result of this paper is that there are a number of both theoretical and phenomenological motivations for considering $\alpha$ close to zero to be the optimal value. These include:
\begin{itemize}
  \item The four dimensional effective Higgs potential receives an additional positive $|\Phi|^2$ term proportional to $\alpha$ which can, for large values of $\alpha$, dominate over the negative $|\Phi|^2$ term and result in an EW phase transition no longer occurring.  
  \item If the space is not cut off in the IR, then the Breitenlohner-Freedman bound implies $\alpha\geqslant 0$ \cite{Breitenlohner:1982jf}.
  \item The holographic method is used to demonstrate, for generic geometries and generic Higgs potentials, that the flatter the Higgs VEV (i.e. the smaller the value of $\alpha $) the smaller the contribution to the Peskin-Takeuchi $T$ parameter \cite{Peskin:1991sw}. 
  \item For the RS model, large values of $\alpha $ can give rise to strongly coupled pseudo-scalars and potentially large constraints from pseudo-scalar mediated flavour changing neutral currents (FCNC's).
  \item It is found, for all spaces considered, that the modification to the $W$ and $Z$ profile will be minimal for smaller values of $\alpha$. This will likely result in a reduction in many of the constraints from flavour physics.
  \item Assuming anarchic Yukawa couplings, there is a general trend that suggests that the smaller the value of $\alpha $ the more the SM fermion profiles will be peaked towards the UV tip of the space. This will again result in the reduction of constraints from FCNC's \cite{Cabrer:2011qb, Agashe:2008uz, Archer:2011bk}.
  \item In order for a 4D Higgs operator, $\mathcal{O}$, with a scaling dimension $2+\alpha$, to offer a resolution to the gauge hierarchy problem then the corresponding $\mathcal{O}^\dag\mathcal{O}$ operator must not be relevant, which implies $\alpha\geqslant 0$ \cite{Luty:2004ye}.
\end{itemize} 

The outline of this paper is as follows. In section \ref{sect:Model} the geometry and the Higgs sector is outlined. In section \ref{sect:FermMass} the fermion mass hierarchy is discussed. In section \ref{sect:EWcons}, the EW constraints are computed. In section \ref{sect:PseudoScal} the pseudo-scalars are investigated. In section \ref{sect:flavour} the implications for flavour physics are discussed and we conclude in section \ref{sect:Conc}. The majority of the core equations have been derived, for a generic space, in the appendix. This paper should be considered as an extension of a number of existing studies on bulk Higgs scenarios, including \cite{Davoudiasl:2005uu, Cacciapaglia:2006mz, Dey:2009gf, Vecchi:2010em, Medina:2010mu}.

\section{The Model.} \label{sect:Model}
This paper seeks to investigate the most minimal extension of the SM in which the Higgs propagates in the bulk of a 5D warped extra dimension. Hence we shall consider a bulk $\rm{SU}(2)\times\rm{U}(1)$ gauge symmetry and not the extended custodial symmetry \cite{Agashe:2003zs}. The fermion content remains the same as the SM and we include a bulk Higgs, $\Phi$, which is a doublet under the $\rm{SU}(2)$ symmetry. As in the RS model, we compactify the space over a $\rm{S}^1/\rm{Z_2}$ orbifold in which the space is cut-off at the two fixed points, $r_{\rm{UV}}=0$ and $r_{\rm{IR}}=R$. The Higgs sector is then described by 
\begin{equation}
\label{ }
S=\int d^5x \sqrt{G}\left [ |D_M\Phi|^2-V(\Phi)\right ]+\int d^4x\sqrt{g_{\rm{IR}}}\left [-V_{\rm{IR}}(\Phi)\right ]_{r=R}+\int d^4x\sqrt{g_{\rm{UV}}}\left [-V_{\rm{UV}}(\Phi)\right ]_{r=0}
\end{equation} 
where $V_{\rm{IR}/\rm{UV}}$ are the Higgs potentials localised on flat branes located at the orbifold fixed points. While $G$ and $g_{\rm{IR}/\rm{UV}}$ are the determinant of the bulk metric and the induced brane metrics. At this point we must make some simplifying assumptions. Firstly we have neglected brane localised kinetic terms and secondly here we consider potentials of the form 
\begin{equation}
\label{ AssumPot}
V(\Phi)=M_\Phi^2|\Phi|^2\qquad V_{\rm{IR}}(\Phi)=-M_{\rm{IR}}|\Phi|^2+\lambda_{\rm{IR}}|\Phi|^4\qquad V_{\rm{UV}}(\Phi)=M_{\rm{UV}}|\Phi|^2.
\end{equation}
We have not included $|\Phi |^4$ terms in the bulk or on the UV brane. In RS type scenarios, the fundamental coefficients of such operators are assumed to be at the Planck scale $\sim\mathcal{O}(k^{-2})$. However, in the 4D effective theory, $\lambda_{\rm{IR}}$ would be warped down to the Kaluza-Klein (KK) scale $\sim\mathcal{O}(M_{\rm{KK}}^{-2})$ while the corresponding UV operator would remain at the Planck scale. A bulk $|\Phi |^4$ term is assumed to be suppressed by an intermediate scale. Although including such a term would result in the Higgs VEV being the solution of a nonlinear differential equation and so result in a significantly more complicated model.   

\subsection{The Geometry.}
With these assumptions the model is largely completely defined and the only remaining input is the geometry of the space. Here we shall consider spaces of the form
\begin{equation}
\label{MetricUsed}
ds^2=e^{-2A(r)}\eta^{\mu\nu}dx_\mu dx_\nu-dr^2
\end{equation}
where $\eta^{\mu\nu}=\rm{diag}(1,-1,-1,-1)$ and $0\leqslant r \leqslant R$. In the original RS model just gravity propagated in the bulk
and the solution considered was a slice of AdS${}_5$,
\begin{equation}
\label{AdSMetric}
A(r)=kr.
\end{equation}
However it was quickly realised that in order to stabilise the space one must also include an additional bulk Goldberger-Wise scalar \cite{Goldberger:1999uk}. Also, in order to generate the fermion mass hierarchy and suppress EW and flavour constraints, the model was extended to allow the SM particles to propagate in the bulk \cite{Grossman:1999ra, Gherghetta:2000qt, Huber:2000ie}. The back reaction of such bulk fields will typically lead to a deviation, from the AdS geometry, in the IR tip of the space. Although the space should be asymptotically AdS in the UV\footnote{The 5D generalisation of Birkhoff's theorem \cite{Bowcock:2000cq} ensures that solutions with no (or constant) bulk fields must be AdS. In the RS model the only fields peaked towards the UV are the light fermion modes which are assumed to have a negligible coupling to gravity. The other fields are typically vanishing in the UV. Hence it is anticipated that the space should be asymptotically AdS in the UV.}.  While it is not certain how large this IR deformation should be, it is important to ask how sensitive any result, based on AdS${}_5$, is to such modifications in the geometry. So in this paper, we shall also consider a class of geometries that have arisen from solving a scalar plus gravity model, giving solutions of the form \cite{Cabrer:2009we, Cabrer:2011fb},  
\begin{equation}
\label{CGQMetric}
A(r)=kr+\frac{1}{v^2}\ln\left (1-\frac{r}{R+\Delta}\right ).
\end{equation}
Firstly, it should be noted that scalar gravity systems generically give rise to a singularity \cite{Gubser:2000nd}, here located at $r=R+\Delta$. While it is possible to construct models that incorporate this singularity, such as soft-wall models, here we shall impose an IR cut-off, on the space, before it reaches the singularity (i.e. let $r\in[0,R]$ and require $\Delta>0$). The parameters $v$ and $\Delta$ are determined by the scalar potential and boundary conditions, although here we shall treat them as free parameters. Note AdS${}_5$ is regained by sending $v$ and $\Delta$ to infinity.

At this point it is useful to define the warp factor and KK scale to be
\begin{equation}
\label{ }
\Omega\equiv e^{A(R)}\hspace{0.8cm}\mbox{and}\hspace{0.8cm}M_{\rm{KK}}\equiv \frac{\partial_5A(r)|_{r=R}}{\Omega}.
\end{equation}
In order for the space to offer the potential for a non-supersymmetric resolution to the gauge hierarchy problem, it is required that the space can be stabilised, such that $\Omega\sim 10^{15}$, and also that $M_{\rm{KK}}\sim\mathcal{O}\,(\rm{TeV})$ is phenomenologically viable.  

\subsection{The Higgs VEV.}\label{sect:HiggsVEV}
In the above model, a bulk Higgs would gain a VEV, $\langle \Phi \rangle=\frac{1}{\sqrt{2}}\left(\begin{array}{c}0 \\h(r)\end{array}\right)$, where $h(r)$ satisfies
\begin{equation}
\label{HiggsVEVode}
\partial_5^2h(r)-4\,\partial_5A(r)\,\partial_5h(r)-M_\Phi^2h=0,
\end{equation}
with the consistent boundary conditions being either $h|_{r=0,R}=0$ or
\begin{equation}
\label{HiggVEVBCs}
\left [\partial_5h-M_{\rm{UV}}h\right ]_{r=0}=0\hspace{0.8cm}\mbox{and}\hspace{0.8cm}\left [\partial_5h-M_{\rm{IR}}h+\lambda_{\rm{IR}}h^3\right ]_{r=R}=0.
\end{equation}
Since we are not considering an EW symmetry breaking bulk potential we must choose the latter `non-Dirichlet' boundary conditions. In other words here EW symmetry is broken on the IR brane. In the early days of the RS model, it was argued that models with a bulk Higgs required fine tuning in order to achieve the correct W and Z masses \cite{Chang:1999nh}. However this work assumed a Higgs VEV that was constant with respect to $r$. From the above relations, it can be seen that a constant Higgs VEV can only be achieved by either breaking EW symmetry on both the branes and the bulk, such that $\partial_\Phi V(\Phi)|_{\Phi=h}=\partial_\Phi V_{\rm{IR}}(\Phi)|_{\Phi=h}=-\partial_\Phi V_{\rm{UV}}(\Phi)|_{\Phi=h}=0$, or alternatively breaking EW symmetry just in the bulk and forbidding the existence of brane potentials. Clearly the first option is finely tuned and it is not clear how the second option could be achieved. Hence here we would suggest that a Higgs VEV peaked towards the IR is a more natural scenario, for which the arguments of \cite{Chang:1999nh} are not applicable.

For the RS model (\ref{AdSMetric}), it is then straight forward to solve for the Higgs VEV (see for example \cite{Luty:2004ye, Cacciapaglia:2006mz})
\begin{equation}
\label{RSHiggsVEV}
h(r)=N_he^{2kr}\left (e^{\alpha kr}+Be^{-\alpha kr}\right )
\end{equation}
where $B$ and $N_h$ are constants of integration fixed by the boundary conditions\footnote{In the special case when $\alpha=0$ then $h(r)=N_h(e^{2kr}+r(M_{\rm{UV}}-2k)e^{2kr})$ and 
\begin{displaymath}
N_h^2=\frac{M_{\rm{UV}}(1+2\ln \Omega)+M_{\rm{IR}}(1-2\ln \Omega)-4k\ln\Omega+RM_{\rm{UV}}M_{\rm{IR}}}{\Omega^4\lambda_{\rm{IR}}(1+R(M_{\rm{UV}}-2k))^3}.
\end{displaymath}},
\begin{equation}
\label{ }
B=-\frac{2k+\alpha k-M_{\rm{UV}}}{2k-\alpha k-M_{\rm{UV}}} \hspace{0.8cm}\mbox{and}\hspace{0.8cm}N_h^2=-\frac{(2k+\alpha k-M_{\rm{IR}})\Omega^{2+\alpha}+B(2k-\alpha k-M_{\rm{IR}})\Omega^{2-\alpha}}{\lambda_{\rm{IR}}\,(\Omega^{2+\alpha}+B\Omega^{2-\alpha})^3}.
\end{equation}
We have also introduced the Higgs exponent, which will prove to be an important parameter,
\begin{equation}
\label{ }
\alpha\equiv\frac{\sqrt{4k^2+M_\Phi^2}}{k}.
\end{equation}
Clearly, one can always take the limit $\alpha\rightarrow\infty$ in order to gain an IR brane localised Higgs. As shall be demonstrated in the next section, in order to explain the observed fermion mass hierarchy it is required that $\alpha$ is small, $\alpha\sim 0$, which clearly requires that $M_\Phi^2<0$. In AdS space, without an IR cut off, a negative mass squared term is permitted provided it satisfied the Breitenlohner-Freedman bound \cite{Breitenlohner:1982jf}. In particular, in order for the total energy of the scalar to be conserved (as well as to allow for a valid global Cauchy surface) it is necessary for the energy-momentum flux to vanish at the AdS boundary. For AdS${}_5$, this implies $M_\Phi^2\geqslant -4k^2$ and $\alpha\geqslant 0$. Technically this bound does not apply here since we are cutting the space off before we reach the AdS boundary. Although the RS model is not a UV complete theory and so it is feasible that such a bound is applicable in a more fundamental realisation of the model. Further still, it was pointed out in \cite{Davoudiasl:2005uu} that such a tachyonic mass term could naturally come from the Higgs coupling to gravity $\sim\zeta \mathcal{R}|\Phi|^2$, where $\zeta$ is an $\mathcal{O}(1)$ coupling and the Ricci scalar is $\mathcal{R}=-20k^2$. 

While this paper will primarily focus on the 5D theory, it is worth making a brief comment about the conjectured dual theory. Although we refer readers to \cite{Luty:2004ye, ArkaniHamed:2000ds, Rattazzi:2000hs, PerezVictoria:2001pa, Batell:2007ez, Gherghetta:2010cj} for a more comprehensive discussion. A bulk Higgs in the RS model is conjectured to be dual to an operator, $\mathcal{O}$, of a broken conformal field theory mixing with an elementary source field $\phi_0=\Phi|_{r=0}$. In other words it is conjectured to be dual to a partially composite Higgs. By computing the bulk-brane propagator and taking the large Euclidean momentum limit, it is found that the scaling dimension of the Higgs operator is given by $2+\alpha$ \cite{Luty:2004ye,Batell:2007ez}. Further still it is found that, below the cut-off $\Lambda\sim k$, the coupling between the source field and operator is approximately $\mathcal{L}_{4D}\sim \frac{\omega}{\Lambda^{\alpha-1}}\phi_0\mathcal{O}$, where $\omega$ is a dimensionless constant. Hence the coupling is relevant for $\alpha<1$ and therefore $\alpha$ determines the level of mixing between the source field and the composite operator. The point we wish to emphasize here is that the parameter $\alpha$ determines both the scaling dimension of the Higgs operator and its level of compositeness, i.e. the larger $\alpha$ the more composite the Higgs. A consequence of this result, discussed in \cite{Luty:2004ye}, is that in order to resolve the hierarchy problem, one require that the corresponding 4D Higgs mass operator, $\mathcal{O}^\dag\mathcal{O}$, is not relevant. This again implies that $\alpha\geqslant 0$.   

The statements of the previous paragraph applies to AdS${}_5$ and while one would anticipate that they should approximately hold when one modifies the geometry, due to the lack of an analytical expression for the bulk-brane propagator, this is challenging to verify. So for the remainder of this paper we shall focus purely on the 5D theory. Turning now to the modified metric (\ref{CGQMetric}), the Higgs VEV can again be found by solving (\ref{HiggsVEVode}) with (\ref{HiggVEVBCs}) to get
\begin{eqnarray}
\label{ }
h(r)=N_he^{(2+\alpha)kr}(R+\Delta-r)^{\frac{v^2-4}{v^2}}\Bigg (U\left (\frac{(v^2-2)\alpha+4}{\alpha v^2},\frac{2v^2-4}{v^2},2\alpha k (R+\Delta-r)\right )\hspace{2cm}\nonumber\\+B\,M\left (\frac{(v^2-2)\alpha+4}{\alpha v^2},\frac{2v^2-4}{v^2},2\alpha k (R+\Delta-r)\right )\Bigg ),
\end{eqnarray}
where $U$ and $M$ are confluent hypergeometric functions or Kummer functions and 
\begin{equation}
\label{ }
\small
B=-\frac{\left ((2k-\alpha k-M_{\rm{UV}})(R+\Delta)+\frac{2\alpha-4}{\alpha v^2}\right )U\left (\frac{(v^2-2)\alpha+4}{\alpha v^2},\frac{2v^2-4}{v^2},2\alpha k(R+\Delta)\right )+U\left (\frac{4-2\alpha}{\alpha v^2},\frac{2v^2-4}{v^2},2\alpha k(R+\Delta)\right )}{\left ((2k-\alpha k-M_{\rm{UV}})(R+\Delta)+\frac{2\alpha-4}{\alpha v^2}\right )M\left (\frac{(v^2-2)\alpha+4}{\alpha v^2},\frac{2v^2-4}{v^2},2\alpha k(R+\Delta)\right )+\frac{4-\alpha v^2+2\alpha}{\alpha v^2}M\left (\frac{4-2\alpha}{\alpha v^2},\frac{2v^2-4}{v^2},2\alpha k(R+\Delta)\right )}.
\end{equation}
Bearing in mind that for large $z>0$, $M(a,b,z)\sim e^zz^{a-b}(1+\mathcal{O}(z^{-1}))$ and $U(a,b,z)\sim z^{-a}(1+\mathcal{O}(z^{-1}))$, then far from the IR tip of the space 
\begin{displaymath}
h(r)\sim N_he^{2kr}\left ((R+\Delta-r)^{-\frac{2(\alpha+2)}{\alpha v^2}}e^{\alpha kr}+B\,(R+\Delta-r)^{-\frac{2(\alpha-2)}{\alpha v^2}}e^{-\alpha kr} \right ).
\end{displaymath}
In other words, in the UV tip of the space one just finds a small power law correction to (\ref{RSHiggsVEV}). However, this approximation breaks down in the IR, particularly for $v\lesssim 1$ when $\frac{(v^2-2)\alpha+4}{\alpha v^2}$ is large. This is a potentially quite interesting region of parameter space in which the Higgs VEV typically grows more than exponentially towards the IR. Unfortunately this greater than exponential growth makes carrying out the numerical studies, conducted later in this paper, quite challenging. So we leave a thorough study of this region of parameter space to future work. Even when $v\gtrsim 1$, the above approximation is still not very robust in the IR. In practice we find a better approximation is
\begin{equation}
\label{ }
h(r)\approx N_he^{(2+\alpha)kr}\left (R+\Delta-r\right )^{\frac{v^2-4}{2v^2}}K_{\frac{v^2-4}{v^2}}\left (2\sqrt{\frac{2k((v^2-2)\alpha+4)(R+\Delta-r)}{v^2}}\right )
\end{equation}  
where $K$ is a modified Bessel function. Note in practice it is found that $B$ is negligibly small for much of the parameter space. This raises the important point that, since we are breaking EW symmetry in the IR and the Higgs VEV is peaked towards the IR, the majority of results of this paper are relatively insensitive to the value of $M_{\rm{UV}}$. This further supports one of our initial assumptions, notably the exclusion of a $|\Phi |^4$ term on the UV brane. For the remainder of this paper we shall fix $M_{\rm{UV}}=k$.  

\subsection{Fixing Some of the Input Parameters.}\label{sect:fixParam}
Even with this relative insensitivity to the UV potential, one may be concerned with the number of free parameters introduced. In particular, in addition to the warp factor and KK scale, the modified metrics have the additional geometrical input parameters $v$ (the IR modification to the curvature) and $\Delta$ (the position of the singularity). While a bulk Higgs, with the assumed potentials (\ref{ AssumPot}), gives rise to an additional four free parameters. Two of these parameters are determined by fitting to the EW scale and the Higgs mass. This leaves four free parameters in addition to those of the RS model with a brane localised Higgs. Here we would argue that this enlargement of the parameter space has arisen from relaxing assumptions of the RS model that (from a bottom up perspective) are questionable. 

We will now move on to fix the two parameters $M_{\rm{IR}}$ and $\lambda_{IR}$. Throughout this paper we shall make the working assumption that the Higgs Mass is 125 GeV. In practice, provided that the Higgs is lighter than the KK scale, changing the Higgs mass will not significantly change the results. The SM Higgs is taken to be the lowest mass eigenstate of (\ref{HiggsPartODE}) with the boundary conditions
\begin{equation}
\label{ HiggsPartBCs}
\left [\partial_5H-M_{\rm{UV}}(h+H)\right ]_{r=0}=0\hspace{0.8cm}\mbox{and}\hspace{0.8cm} \left [\partial_5 H-M_{\rm{IR}}(h+H)+\lambda_{\rm{IR}}\left (\frac{3}{2}h^2H+h^3\right )\right ]_{r=R}=0.
\end{equation}
where $h$ and $H$ are defined in (\ref{PHIDEF}). Of course one could also impose Dirichlet boundary conditions (DBC's) on the Higgs. This would significantly change the phenomenological implications of this model and would, for example, result in no light Higgs. However in light of recent LHC results we consider this scenario disfavoured. Also note that if the Higgs was just localised on the IR brane then the first `$\partial_5$' terms in (\ref{HiggVEVBCs}) and (\ref{ HiggsPartBCs}) would not be present. This would imply the familiar relation $h=\left(\frac{M_{\rm{IR}}}{\lambda_{\rm{IR}}}\right )^{\frac{1}{2}}$ and result in the tree level Higgs mass being given by just $M_{\rm{IR}}$.   

For the RS model, (\ref{HiggsPartODE}) can be solve analytically to get
\begin{equation}
\label{HiggsPartProfile}
f_n^{(H)}=N_He^{2kr}\left (J_\alpha\left (\frac{m^{(H)}_ne^{kr}}{k}\right )+\beta Y_\alpha\left (\frac{m^{(H)}_ne^{kr}}{k}\right )\right )
\end{equation}
where we have made the usual KK decomposition, $H(x,r)=\sum_n f_n^{(H)}(r)H^{(n)}(x)$, such that $\partial_\mu\partial^\mu H^{(n)}=-m_n^{(H)\,2}H^{(n)}$. While $J$ and $Y$ are Bessel functions and $\beta$ is a constant of integration fixed by the boundary conditions. When considering the modified metrics one must work with a numerical solution for the Higgs profile. In addition to the Higgs mass we also fit to the Fermi constant, $\hat{G}_f=1.166367(5)\times 10^{-5}$, the Z mass, $\hat{M}_Z=91.1876\pm0.0021$ GeV and the fine structure constant, $\hat{\alpha}(M_Z)^{-1}=127.916\pm0.0015$ \cite{Nakamura:2010zzi}. These are given, at tree level, by
\begin{eqnarray}
\hat{M}_Z=m_0^{(Z)} \hspace{0.8cm}
\sqrt{4\pi \hat{\alpha}(M_Z)}=\frac{gg^{\prime}}{\sqrt{g^2+g^{\prime\,2}}}f_0^{(\gamma)} \hspace{0.8cm}\nonumber\\
4\sqrt{2} \hat{G}_f=g^2\sum_n\frac{\left (\int dr\, e^Af_0^{(\mu_L)}f_n^{(W)}f_0^{(\nu_{\mu \,L})}\right )\;\left (\int dr\, e^Af_0^{(e_L)}f_n^{(W)}f_0^{(\nu_{e\,L})}\right )}{m_n^{(W)\,2}}\label{EWobserv}
\end{eqnarray}
where $f_n^{(\gamma,W)}$ are the photon and W profiles, defined in the appendix. While $f_0^{(\mu_L,\nu_{\mu\,L},e_L,\nu_{e\,L})}$ are the fermion zero mode profiles for fermions with a bulk mass parameter $c_L$, given in (\ref{fermProf}). For the purpose of this fit, we assume that the muon, electron and neutrinoes all have a universal bulk position $c_L=-c_R=0.7$. Also note that, with the absence of a right handed neutrino zero mode, the charged pseudo-scalars will not contribute at tree level to the Fermi constant (see section \ref{sect:PseudoScal}).  

\begin{figure}[ht!]
    \begin{center}
        \subfigure[The RS Model]{%
            \label{fig:MIRLIRRS}
            \includegraphics[width=0.49\textwidth]{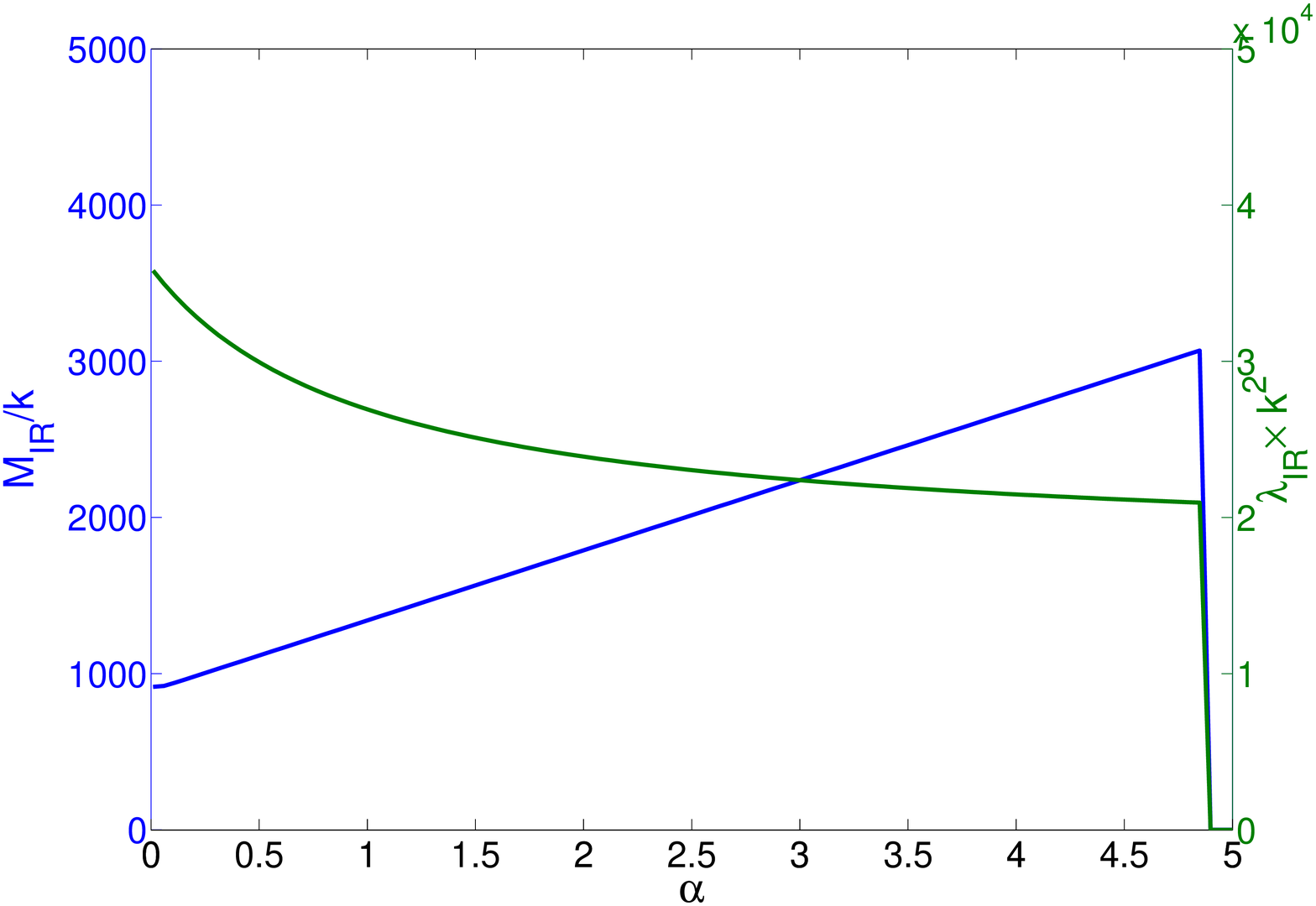}
        }\\
        \subfigure[Modified metric with $v=10$]{%
           \label{fig:second}
           \hspace{-1.5cm}\includegraphics[width=0.49\textwidth]{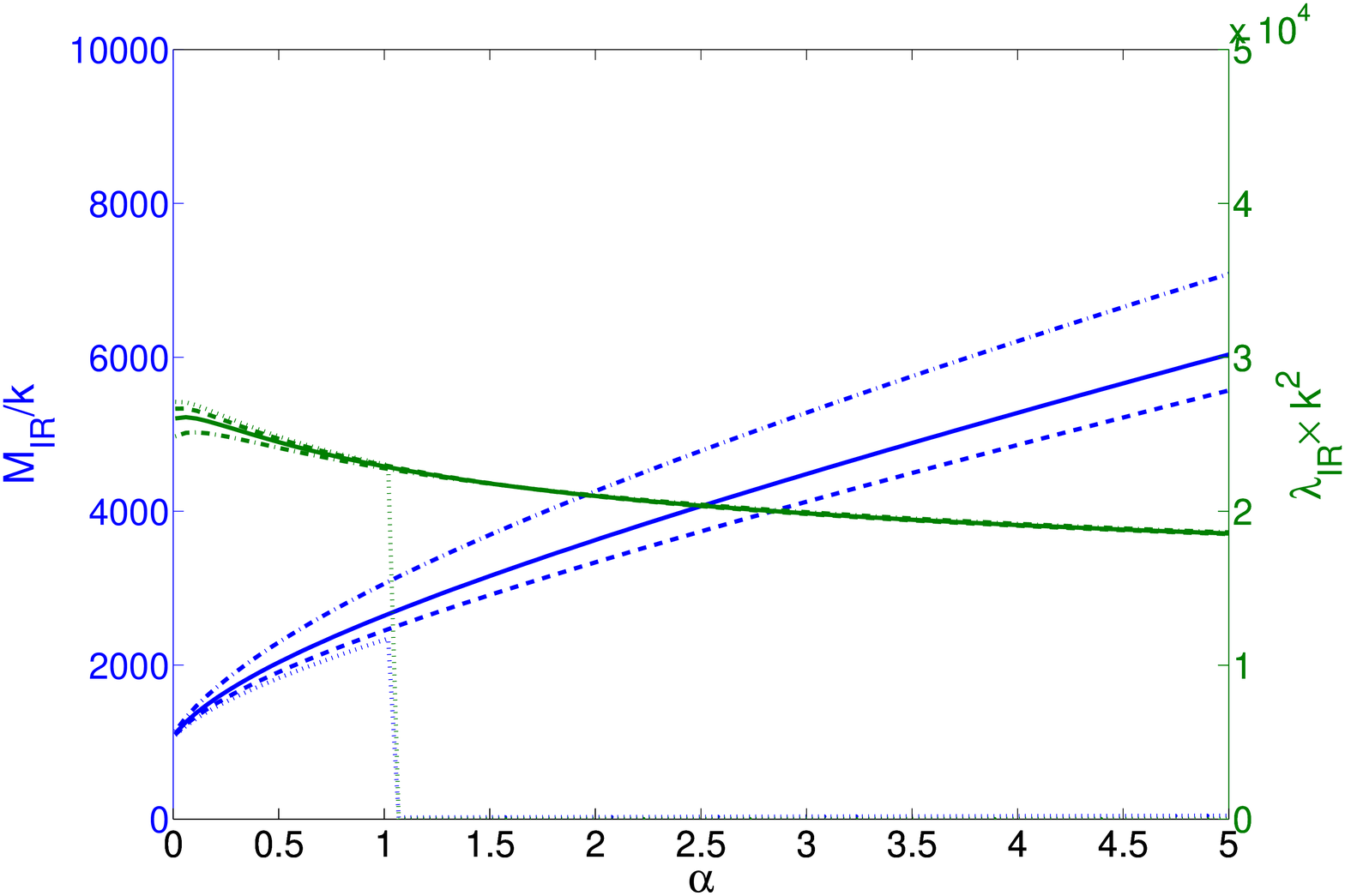}
        }
        \subfigure[Modified metric with $v=3$]{%
            \label{fig:FermMass2}
            \includegraphics[width=0.51\textwidth]{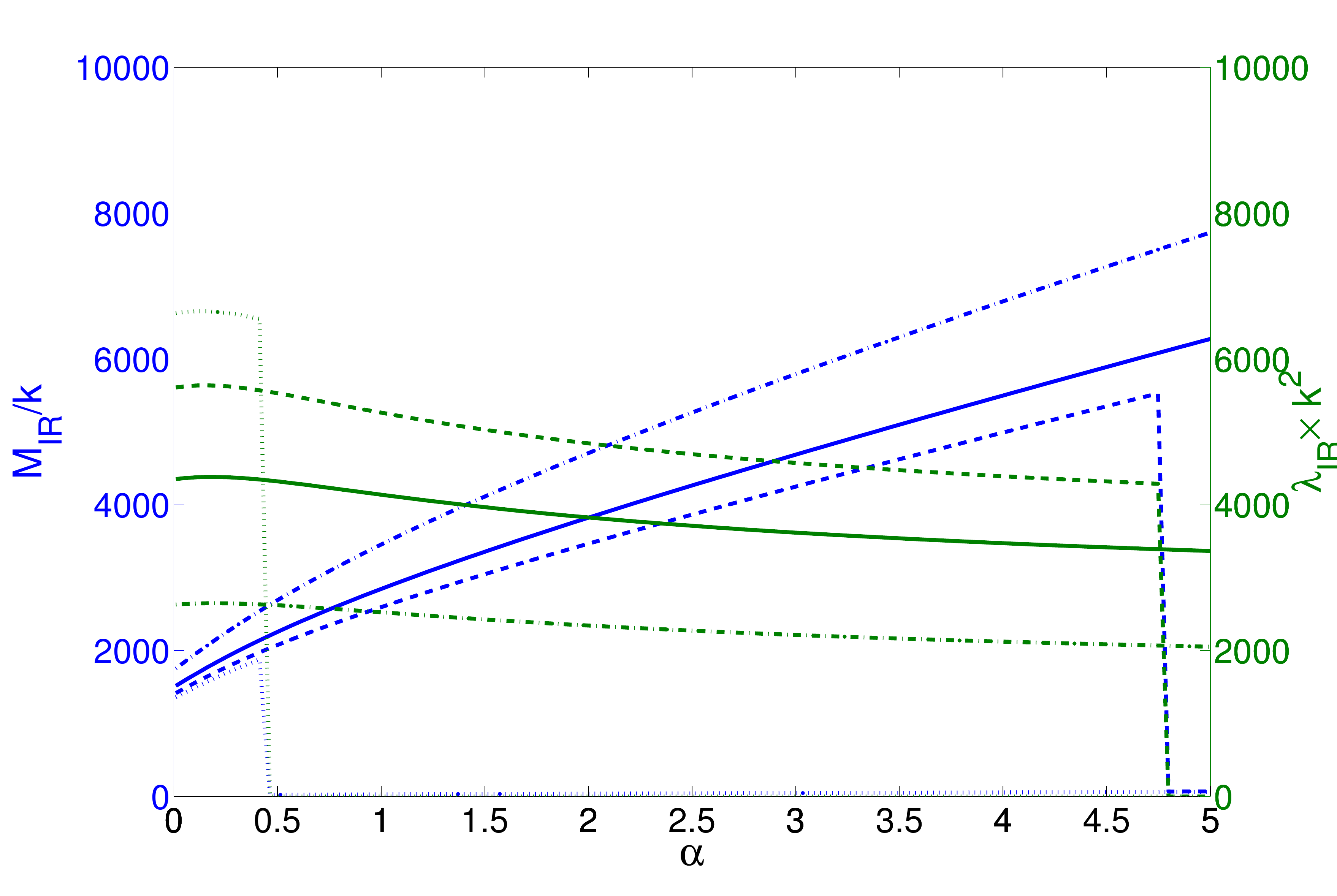}
        }%   \label{fermmass}
    \end{center}
    \caption{The values of $M_{\rm{IR}}$ and $\Lambda_{\rm{IR}}$ after fitting to $\hat{M}_Z$, $\hat{\alpha}$, $\hat{G}_f$ and $m_0^{(H)}=125$ GeV. For the modified metrics we have taken $k\Delta=0.5$ (dash-dot), $k\Delta=1$ (solid),  $k\Delta=1.5$ (dash-dash) and $k\Delta=2$ (dots). We have fixed $\Omega=10^{15}$, $M_{\rm{UV}}=k$ and $M_{\rm{KK}}=2$ TeV.} \label{fig:LambIRMIR}
\end{figure}

The results have been plotted in figure \ref{fig:LambIRMIR}. This brings us to the first motivation for favouring small values of $\alpha$.  In the standard Higgs mechanism one requires a negative $|\Phi|^2$ term, in the Higgs potential, in order to generate a non zero Higgs VEV and break EW symmetry. When one allows the Higgs to propagate in the bulk of an extra dimension, the 4D effective potential will gain, at tree level, a positive contribution to the $|\Phi|^2$ term loosely related to $\partial_5 \Phi$. The larger the value of $\alpha$, the larger this contribution will be and if $\alpha$ is too large then EW symmetry will no longer be broken at tree level. This can be seen explicitly in figure \ref{fig:LambIRMIR}. As $\alpha$ is increased, the require negative $M_{\rm{IR}}$ also increases, in order to compensate the positive contribution and at some point ($\alpha \sim 5$ for the RS model) the positive contribution dominates and EW symmetry is no longer broken.   

One should be cautious about taking the values of $\alpha$ in figure \ref{fig:LambIRMIR}, for which this happens, too literally. Where it is not possible to find values of $M_{\rm{IR}}$ and $\lambda_{\rm{IR}}$ that break EW symmetry, one should of course consider additional operators, in particular a bulk $|\Phi |^4$ term and it is not clear to what extent this effect continues beyond tree level. Rather figure \ref{fig:LambIRMIR} gives an indicator of the values of $\alpha$ for which breaking EW symmetry, in a fashion that is compatible with EW observables, becomes difficult. What is surprising is that for certain geometries, in particular geometries with small $v$ and large $\Delta$, this can happen for relatively small values of $\alpha$.      

\section{Fermion Masses.}\label{sect:FermMass}

Having found the form of the Higgs VEV, we can now move on to look at the fermion masses. The fermions will gain a mass via the Yukawa coupling, $Y$, to the Higgs. I.e. in addition to the terms in (\ref{FermAction}) the action will also include 
\begin{displaymath}
\mathcal{L}_{\rm{Yukawa}}=-Y_D\bar{\Psi}\Phi X-Y_U\epsilon^{ab}\bar{\Psi}_a\Phi_b^\dag X
\end{displaymath}
where $\Psi$ is a doublet under SU($2$) and $X$ is a singlet. If we split the spinor into its chiral components $X=\chi_L+\chi_R$ and likewise for $\Psi$  (as in the appendix) then, after compactifying over the $S^1/Z_2$ and choosing appropriate boundary conditions, only $\psi_L$ and $\chi_R$ will gain zero modes. So a low energy chiral theory is achieved with $\Psi\ni\{L,Q\}$ and $X\ni\{e,u,d\}$.  It can now be seen that, due to this mixing by the bulk Higgs, the profiles given in (\ref{FermEOM}) will not be mass eigenstates. To compute the mass eigenstates, if we define
\begin{equation}
\label{ }
Y_{LR}^{(n,m)}\equiv \frac{1}{\sqrt{2}}Y\int dr\;  hf_n^{(\psi_{L})}f_m^{(\chi_{R})}\qquad\mbox{ and }\qquad Y_{RL}^{(n,m)}\equiv \frac{1}{\sqrt{2}}Y\int dr\;  hf_n^{(\psi_{R})}f_m^{(\chi_{L})}
\end{equation}  
then the fermion mass matrix will be given by
\begin{equation}
\label{FermionMassMatrix}
\left(\begin{array}{cccc}\bar{\psi}_L^{(0)} & \bar{\psi}_L^{(1)} & \bar{\chi}_L^{(1)} & \dots\end{array}\right)\left(\begin{array}{ccccc}Y_{LR}^{(0,0)} & 0 & Y_{LR}^{(0,1)} & 0 & \dots \\Y_{LR}^{(1,0)} & m_1^{(\psi)} & Y_{LR}^{(1,1)} & 0 &  \\0 & Y_{RL}^{(1,1)} & m_1^{(\chi)} & Y_{RL}^{(2,1)} &  \\Y_{LR}^{(2,0)} & 0 & Y_{LR}^{(2,1)} & m_2^{(\psi)} &  \\\vdots &  &  &  & \ddots\end{array}\right)\left(\begin{array}{c}\chi_R^{(0)} \\\psi_R^{(1)} \\\chi_R^{(1)} \\\psi_R^{(2)} \\\vdots\end{array}\right).
\end{equation}
Note we have restricted this discussion to one generation, in reality $Y_{LR}$ and $Y_{RL}$ will be block $3\times 3$ matrices. This matrix can then be perturbatively diagonalised (see for example \cite{Archer:2010hh}) such that the zero mode mass eigenvalue is found to be approximately 
\begin{equation}
\label{ }
\tilde{m}_0\approx Y_{LR}^{(0,0)}+\sum_{n,m=1}\frac{Y_{LR}^{(0,n)}Y_{RL}^{(n,m)}Y_{LR}^{(m,0)}}{(m_n^{(\chi)}-Y_{LR}^{(0,0)})(m_m^{(\psi)}-Y_{LR}^{(0,0)})}+\mathcal{O}(m_n^{-3}).
\end{equation}     
While the profile of the mass eigenstate of the $\chi_R$ zero mode, for example, will be
\begin{equation}
\label{fermCorrections}
\tilde{f}_0^{(\chi_R)}\approx f_0^{(\chi_R)}-\sum_{n=1}\frac{Y_{LR}^{(0,n)}}{(m_n^{(\chi)}-Y_{LR}^{(0,0)})}f_n^{(\chi_R)}+\mathcal{O}(m_n^{-2}).
\end{equation}   
So the SM fermion masses will be given by $Y_{L,R}^{(0,0)}$ with an $\mathcal{O}(\tilde{m}_0^2/ M_{\rm{KK}}^2)$ corrections and will have the profiles $f_0^{(\psi, \chi)}$ with $\mathcal{O}(\tilde {m}_0/ M_{\rm{KK}})$ corrections. In this paper we will be largely concerned with the lighter fermions and so 
we will neglect these corrections. However these corrections should not be completely forgotten since they may lead to phenomenological deviations from the brane Higgs scenario, particularly with regard to top physics. It is also possible that in models with extended Higgs sectors, such as models with a custodial symmetry \cite{Agashe:2003zs} or models with axial gluons \cite{Bauer:2011ah}, these corrections may be enhanced.    

\begin{figure}[t!]
    \begin{center}
        \subfigure[Brane localised Higgs]{%
            \label{fig:BraneHiggs}
            \includegraphics[width=0.48\textwidth]{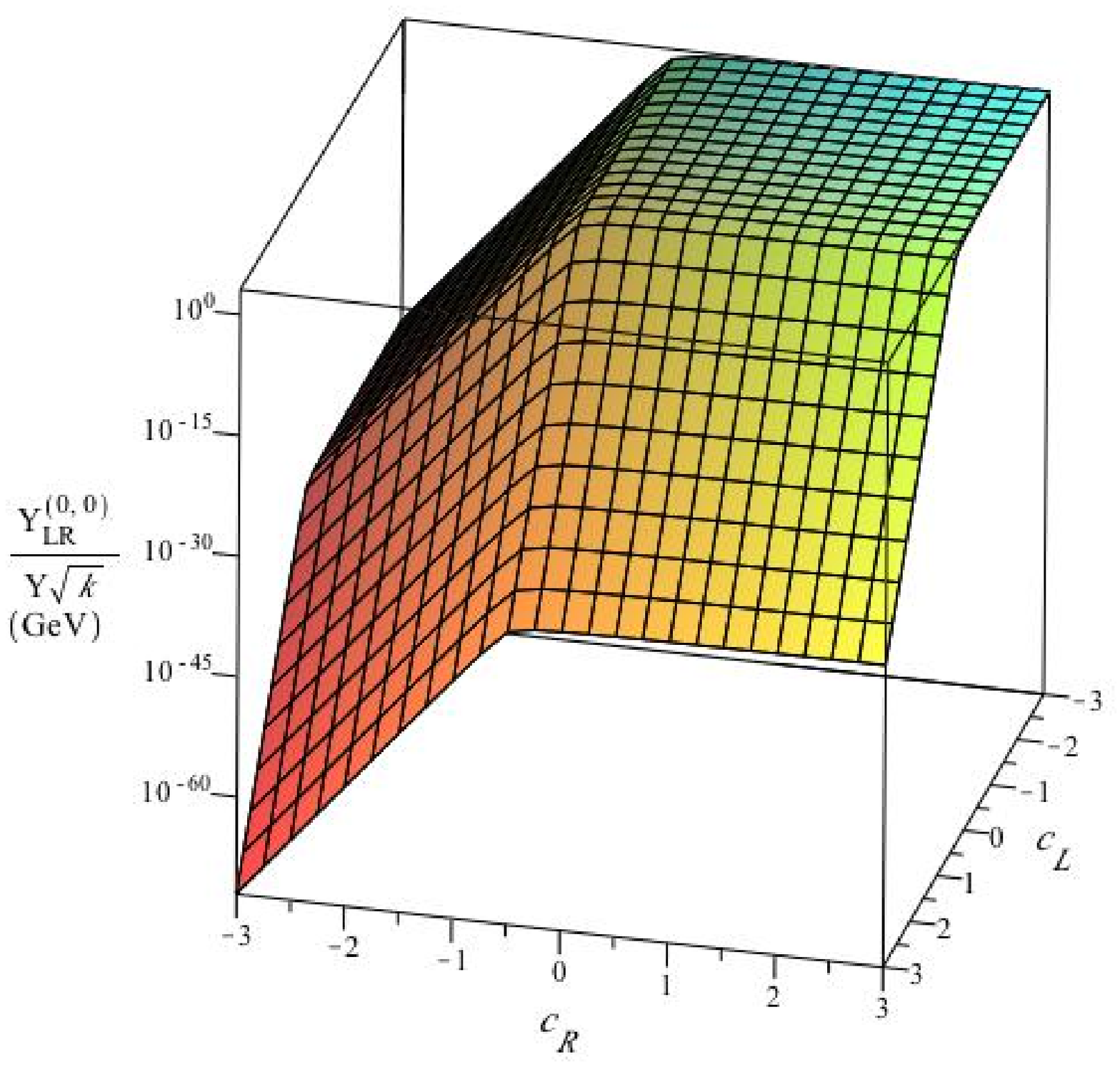}
        }
        \subfigure[Bulk Higgs ($\alpha=0.01$)]{%
           \label{fig:BulkHiggs}
           \includegraphics[width=0.48\textwidth]{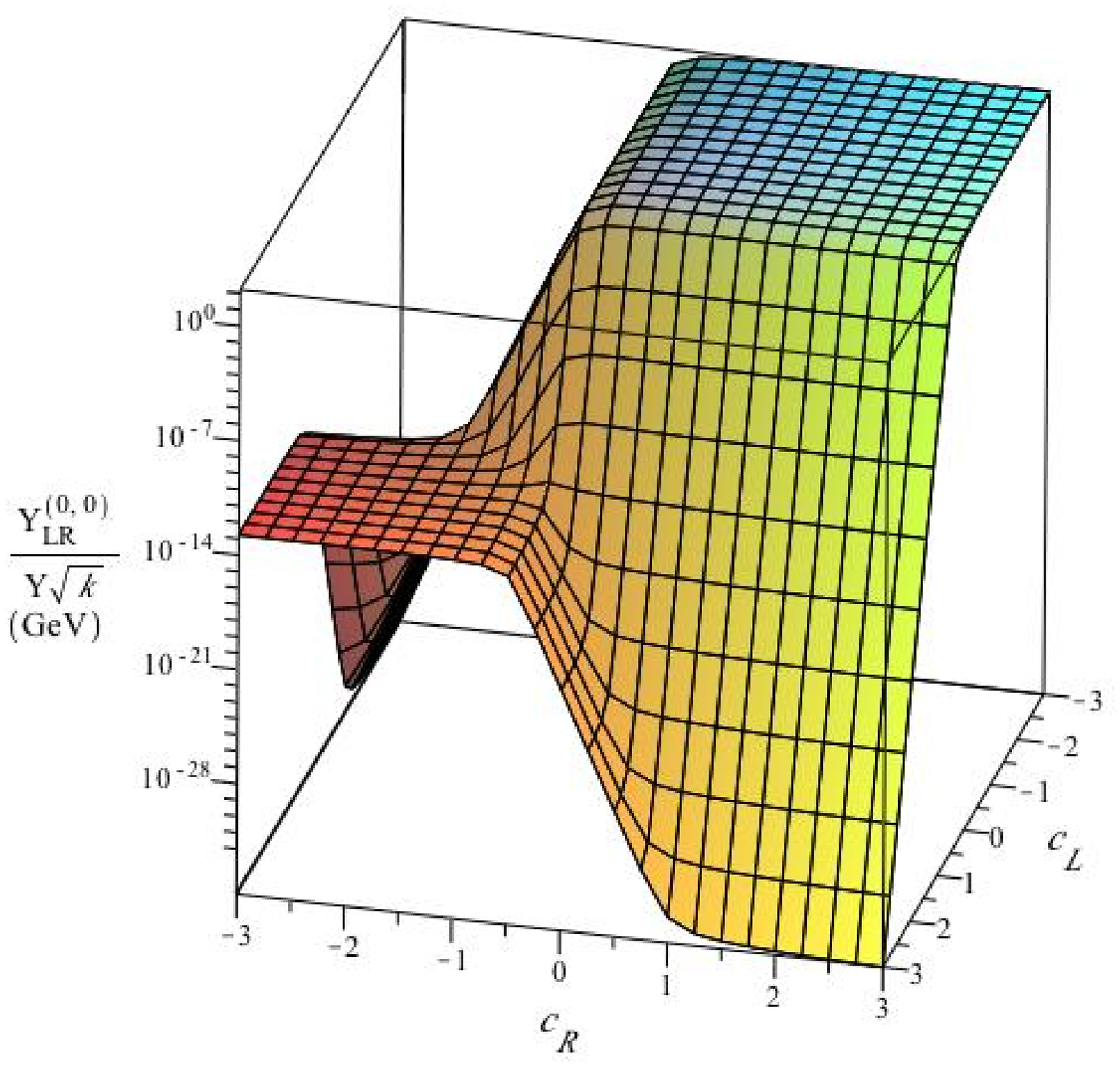}
        }
    \end{center}
    \caption{The approximate range of fermion zero mode masses, $\frac{Y_{LR}^{(0,0)}}{\tilde{Y}}$, for a bulk and brane Higgs in the RS model. The fermions are peaked towards the UV (IR) when $c_L>0.5$ ( $c_L<0.5$) and $c_R<-0.5$ ($c_R>-0.5$). $\Omega=10^{15}$ and $M_{\rm{KK}}=2$ TeV. } \label{fermmass}
\end{figure}

For the RS model it is then straightforward to analytically integrate over (\ref{RSHiggsVEV}) and the fermion profiles (\ref{fermProf}) giving
\begin{eqnarray}
\label{ }
Y_{LR}^{(0,0)}=\frac{\tilde{Y}}{\sqrt{2}}M_{\rm{KK}}\sqrt{\frac{(1-2c_L)(1+2c_R)\left ((\tilde{M}_{\rm{IR}}-2-\alpha)\Omega^\alpha+B(\tilde{M}_{\rm{IR}}-2+\alpha)\Omega^{-\alpha}\right )}{(\Omega^{1-2c_L}-1)(\Omega^{1+2c_R}-1)\;\tilde\lambda_{IR}\;\left (\Omega^\alpha+B\Omega^{-\alpha}\right )^3}}\times\hspace{1cm}\nonumber\\
\left (\frac{\Omega^{1-c_L+c_R+\alpha}-\Omega^{-1}}{(2-c_L+c_R+\alpha)}+B\frac{\Omega^{1-c_L+c_R-\alpha}-\Omega^{-1}}{(2-c_L+c_R-\alpha)}\right ),\label{RSYLR}
\end{eqnarray}
where we have introduced the `assumed $\mathcal{O}(1)$' coefficients $\tilde{M}_{\rm{IR},\rm{UV}}=M_{\rm{IR},\rm{UV}}k^{-1}$, $\tilde{\lambda}_{\rm{IR}}=\lambda_{\rm{IR}}k^2$ and $\tilde{Y}=Y\sqrt{k}$. Although in practice, since the EW scale is not the same as the KK scale, these parameters are not $\mathcal{O}(1)$ (see figure \ref{fig:LambIRMIR}). This should be contrasted with the analogous expression for the brane localised Higgs (see for example \cite{Huber:2003tu}),
\begin{equation}
\label{ }
Y_{LR}^{(0,0)}=\frac{\tilde{Y}}{\sqrt{2}}v_4\sqrt{\frac{(1-2c_L)(1+2c_R)}{(\Omega^{1-2c_L}-1)(\Omega^{1+2c_R}-1)}}\Omega^{1-c_L+c_R}
\end{equation}
where the 4D Higgs VEV is $v_4\approx 246$ GeV\footnote{In practice $v_4$, in the RS model with a brane localised Higgs, will still differ from the SM value. See \cite{Bouchart:2009vq}.}. The range of fermion masses have been plotted in figure \ref{fermmass}. For the modified metrics, $Y_{L,R}^{(0,0)}$ does not have a neat analytical form but the resulting distribution of fermion masses is similar to that which is plotted in figure \ref{fig:BulkHiggs}.

\begin{figure}[ht!]
    \begin{center}
        \subfigure[`Minimum mass'=$Y_{LR}^{(0,0)}/\tilde{Y}$ with $c_L=5, c_R=-5$]{%
            \label{fig:MinMass}
            \includegraphics[width=0.68\textwidth]{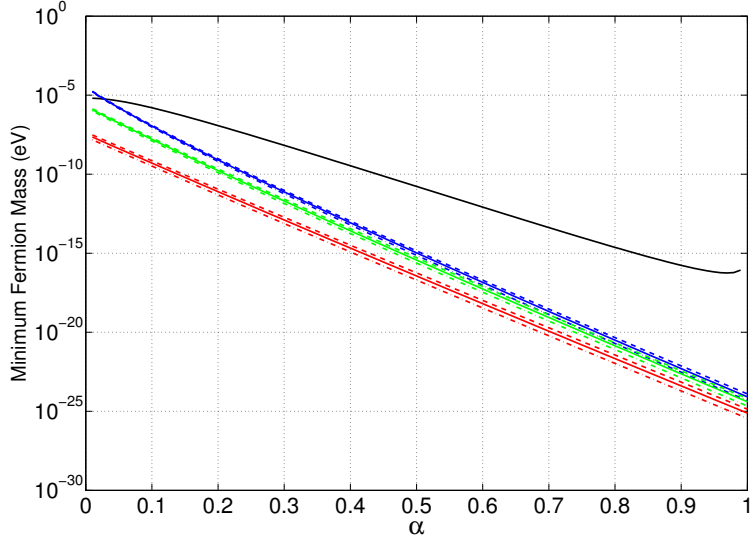}
        }\\
        \subfigure[`Maximum mass'=$Y_{LR}^{(0,0)} /\tilde{Y}$ with $c_L=-3, c_R=3$]{%
           \label{fig:MaxMass}
           \includegraphics[width=0.68\textwidth]{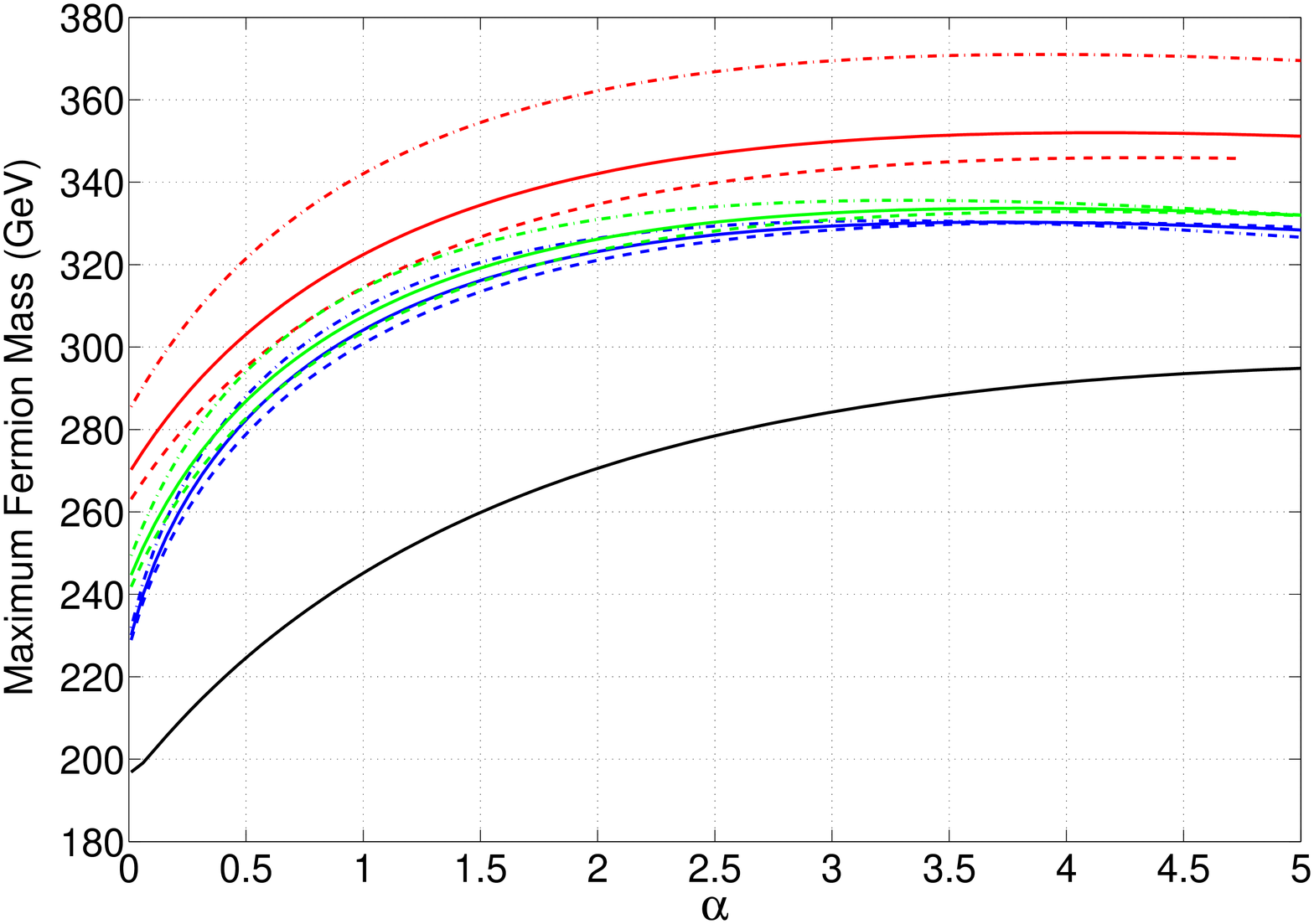}
        }
    \end{center}
    \vspace{-0.5cm}
    \caption{The approximate upper and lower limits of the fermion zero mode mass, assuming $\tilde{Y}\approx 1$ and one does not have a split fermion scenario. Here we have considered the RS model (black) as well as the modified metrics with $v=10$ (blue), $v=5$ (green) and $v=3$ (red) with $k\Delta=1.5$ (dash-dash line), $k\Delta=1$ (solid line) and $k\Delta=0.5$ (dash-dot line). We have set $M_{\rm{UV}}=k$, $M_{\rm{KK}}=2$ TeV and $\Omega=10^{15}$. The remaining input parameters have been fixed as described in section \ref{sect:fixParam}. The `maximum fermion mass' does decrease for large $\alpha$ and it is suspected that it should flow towards $\sim174$ GeV as $\alpha\rightarrow \infty$ for large $M_{\rm{KK}}$. These large values of $\alpha$ have not been included in the plot since they would not give rise to a phenomenologically viable electroweak phases transition, as described in the previous section. For large values of $\alpha$ the lower `plateau' will not have begun for the bulk mass parameters $c_L=3$ and $c_R=-3$ and so the minimum fermion mass is evaluated at $c_L=5$ and $c_R=-5$. } \label{MinMaxMass}
\end{figure}

This brings us to the central result of this paper and the initial motivation for this study. It is already widely known that warped extra dimensions offer a potential explanation for the large hierarchies that exist in the observed fermion masses \cite{Grossman:1999ra, Gherghetta:2000qt, Huber:2000ie}. As can be seen in figure \ref{fig:BraneHiggs}, order one changes in the bulk mass parameter result in exponential changes in the zero mode mass. However the model offers no indication of what the overall size of the mass hierarchy should be. When we consider a bulk Higgs this situation changes. When the fermion profiles are peaked towards the IR brane, the $\Omega^{1-c_L+c_R+\alpha}$ term will dominate and $m_0\sim \tilde{Y}M_{\rm{KK}}\tilde{\lambda}^{-\frac{1}{2}}_{\rm{IR}}\sim 200-300$ GeV. Although when the fermions are peaked towards the UV brane the second $\Omega^{-1}$ term will dominate and $m_0\sim  \tilde{Y}M_{\rm{KK}}\tilde{\lambda}^{-\frac{1}{2}}_{\rm{IR}}\Omega^{-1-\alpha}$. Hence if we assume there are no hierarchies in $\tilde{Y}$ and we do not have a split fermion scenario, then the difference between the lightest and the heaviest fermion Dirac mass term is $\lesssim \Omega^{-1-\alpha}$. 

These are of course approximate relations, but the calculated values are plotted in figure \ref{MinMaxMass}. This figure requires a number of remarks. Firstly, what has been plotted is not the absolute maximum and minimum mass but rather the location of the upper and lower `plateaus' in figure \ref{fig:BulkHiggs}. These plateaus are not strictly flat but increase (or decrease) logarithmically and also the bulk mass parameters, for which the lower `plateau' begins, increases as $\alpha$ is increased (see analogous plots in \cite{Archer:2011bk}). We have also assumed $\tilde{Y}\approx 1$. By naive dimensional analysis it is suspected that perturbative control of the theory is lost if $\tilde{Y}\gtrsim 3$ \cite{Csaki:2008zd}, but there is nothing, except naturalness arguments, forbidding $\tilde{Y}$ from being small. Likewise we have rejected the possibility of a split fermion scenario in which the left handed fermions are peaked towards the UV and the right handed fermions are peaked towards the IR or vice-versa. Such a scenario would undoubtably give rise to large constraints from FCNC's and so are typically disfavoured by fits including relevant flavour observables (see section \ref{sect:flavour}).\newpage    

If we accept these assumptions, then we arrive at a potentially interesting result, first pointed out in \cite{Agashe:2008fe}. For spaces close to AdS${}_5$, with a small value of $\alpha\lesssim 0.1$, the lower bounds on the fermions masses are remarkably close to the suspected value for the lightest neutrino mass, $10^{-4}-10^{-2}$ eV\footnote{While there is considerable uncertainty in this mass, assuming a normal hierarchy, the recent measurement of a large value of $\theta_{13}$ would probably favour a lighter minimum neutrino mass.  See for example \cite{Pascoli:2007qh}}. This alone does not explain the size of the neutrino masses, since the above results are for the Dirac mass term. Hence one must also demonstrate why either the Majorana mass term is forbidden, or alternatively explain why it exists at approximately the same scale. There are many possibilities for achieving this, for example, the latter option could be achieved if the origin of the Majorana mass term had a similar profile to the Higgs VEV.  However these possibilities would necessarily require an extension of the minimal set up considered in this paper and so here we leave them to future work. Also, a bulk Higgs scenario offers no explanation as to why there is a six orders of magnitude `desert' between the electron mass, and the heaviest neutrino mass or why the leptons are lighter than the quarks. In other words, this work is concerned with the range of the Dirac mass term, while statements concerning neutrino masses necessarily require an extended model.   

Having said that, here we would argue that if one could understand why $\alpha$ should be small, then models with a bulk Higgs would offer some insight into why the fermion mass hierarchy is the size that it is. Having already offered two motivations for favouring a small value of $\alpha$, we shall now move on to demonstrate that many of the phenomenological constraints on warped extra dimensions will be at their minimum for smaller values of $\alpha$.     
 
\section{Electroweak Constraints.}\label{sect:EWcons}
One of the first tests of any model, that modifies the Higgs mechanism of the SM, is that of being able to suppress corrections to the precision EW observables. Typically it is found that such corrections are dominated by modifications to the $W$ and $Z$ propagators and are often parameterised in terms of the `oblique'  Peskin-Takeuchi parameters \cite{Peskin:1991sw}. Such a parameterisation assumes that the corrections to gauge-fermion couplings are negligible and the new physics can be described by the effective Lagrangian
\begin{equation}
\label{STeffLang}
\mathcal{L}_{\rm{eff.}}=-\frac{1}{2}Z_\mu\left (p^2-m_Z^2-\Pi_Z(p)\right )Z^\mu-W_\mu^{+}\left (p^2-m_W^2-\Pi_W(p)\right )W^\mu_{-}+\dots
\end{equation}    
where $p=p^\mu$ is the four momentum. The effective Lagrangian also contains terms for the photon as well as terms mixing the photon and the $Z$, but these corrections are found to be zero, at tree level, in models with extra dimensions. Expanding the $\Pi$'s as 
\begin{displaymath}
\Pi(p)=\Pi(0)+p^2\Pi^{\prime}(0)+\frac{p^4}{2}\Pi^{\prime\prime}(0)+\dots
\end{displaymath}  
then the Peskin-Takeuchi $S$ and $T$ parameters are defined to be
\begin{equation}
\label{ }
\hat{\alpha}T=\frac{1}{m_Z^2}\left (\Pi_Z(0)-\frac{1}{\sin^2 \theta_w}\Pi_W(0)\right )\hspace{0.8cm}\mbox{and}\hspace{0.8cm}\hat{\alpha}S=4\sin^2 \theta_w\cos^2 \theta_w\left (\Pi_Z^{\prime}(0)-\Pi_\gamma^\prime(0)\right ),
\end{equation}
where $\theta_w$ is the weak mixing angle. It is possible to introduce additional parameters in order to account for the higher order corrections \cite{Cacciapaglia:2006pk}. However here it is found that, with the absence of a custodial symmetry \cite{Agashe:2003zs}, the constraint is dominated by contributions to the $T$ parameter.

In order to compute the corrections to the propagator here we shall largely follow \cite{Cabrer:2011fb} and expand the 5D field in terms of the holographic basis. This method is completely equivalent to a KK expansion and does not rely on the existence of a dual theory. In the interests of generality, we shall also work with the generic metric considered in the appendix (\ref{genericMetric }). In particular we factorise the 5D fields
\begin{equation}
\label{ }
Z_\mu(p,r)=G^{(Z)}(p,r)\tilde{Z}_\mu(p)\hspace{0.8cm}\mbox{such that}\hspace{0.8cm} Z_\mu(p,r_{\rm{UV}})=\tilde{Z}_\mu(p)
\end{equation}
and likewise for the $W$. We also impose Neumann boundary conditions (NBC's) on just the IR brane, $\partial_5Z_\mu |_{r_{\rm{IR}}}=0$. Bearing in mind that $G^{(Z)}$ will satisfy $\partial_5(a^2b^{-1}\partial_5G^{(Z)})-a^2bM_Z^2G^{(Z)}+bp^2G^{(Z)}=0$, the tree level effective Lagrangian will be given purely by the UV boundary term
\begin{equation}
\label{effLangObli}
\mathcal{L}_{\rm{eff.}}=\frac{1}{2}\tilde{Z}_\mu \left [G^{(Z)}a^2b^{-1}\partial_5G^{(Z)}\right ]_{r=r_{\rm{UV}}}\tilde{Z}^\mu+\tilde{W}^+_\mu \left [G^{(W)}a^2b^{-1}\partial_5G^{(W)}\right ]_{r=r_{\rm{UV}}}\tilde{W}_-^\mu
\end{equation}
We have once again neglected possible brane localised kinetic terms which are known to reduce the EW constraints \cite{Carena:2002dz}. To proceed further we define
\begin{equation}
\label{ }
P_{W,Z}(p,r)\equiv\frac{a^2b^{-1}\partial_5G^{(W,Z)}(p,r)}{G^{(W,Z)}(p,r)}
\end{equation}
such that
\begin{equation}
\label{Peqn}
\partial_5 P_{W,Z}+a^{-2}bP_{W,Z}^2-a^2M_{W,Z}^2+bp^2=0.
\end{equation}
We can now match $P_{W,Z}$ to the oblique parameters by expanding in four momentum
\begin{equation}
\label{ }
P_{W,Z}(p,r)=P_{W,Z}^{(0)}(r)+p^2P_{W,Z}^{(1)}(r)+\frac{1}{2}p^4P_{W,Z}^{(2)}(r)+\dots
\end{equation}
and equating (\ref{STeffLang}) to (\ref{effLangObli}) giving
\begin{equation}
\label{ }
\Pi_{W,Z}(0)=P_{W,Z}^{(0)}\big |_{r=r_{\rm{UV}}}-m_{W,Z}^2\hspace{0.8cm}\mbox{and}\hspace{0.8cm}\Pi^{\prime}_{W,Z}(0)=P_{W,Z}^{(1)}\big |_{r=r_{\rm{UV}}}+1.
\end{equation}
Where $P^{(n)}$ can be found by expanding (\ref{Peqn}) into a set of coupled equations
\begin{eqnarray}
\partial_5P_{W,Z}^{(0)}+a^{-2}bP_{W,Z}^{(0)\,2}-a^2bM_{W,Z}^2=0\nonumber\\
\partial_5P_{W,Z}^{(1)}+2a^{-2}bP_{W,Z}^{(0)}P_{W,Z}^{(1)}+b=0\nonumber\\
\partial_5P^{(2)}_{W,Z}+a^{-2}b(P_{W,Z}^{(0)}P_{W,Z}^{(2)}+P_{W,Z}^{(1)\,2})=0\label{Pdecompose}\\
\vdots\nonumber
\end{eqnarray}
As already mentioned, the greatest contribution to the EW constraints comes from the $T$ parameter which is given by $P_{W,Z}^{(0)}\big |_{r=r_{\rm{UV}}}$. Without loss of generality, we can set $b=1$, $a(r_{\rm{IR}})=\Omega^{-1}$ and $a(r_{\rm{UV}})=1$. Then using the IR boundary condition, $P_{W,Z}(r_{\rm{IR}})=P^{(0)}_{W,Z}(r_{\rm{IR}})=0$, (\ref{Pdecompose}) can be rearranged to give $P_{W,Z}^{(0)}\big |_{r=r_{\rm{UV}}}$ (and hence the $T$ parameter) as
\begin{equation}
\label{P0eval}
P_{W,Z}^{(0)\,2}\big |_{r=r_{\rm{UV}}}=\partial_5P_{W,Z}^{(0)}\big |_{r=r_{\rm{IR}}}-\partial_5P_{W,Z}^{(0)}\big |_{r=r_{\rm{UV}}}+M_{W,Z}^2|_{r=r_{\rm{UV}}}-\Omega^{-2}M_{W,Z}^2|_{r=r_{\rm{IR}}}.
\end{equation}
It should be stressed that this holds for generic Higgs VEV's and generic 5D geometries. This then brings us to the point of this discussion. Bearing in mind the definition of $M_{W,Z}$ (\ref{ MWZdef}), it is now straight forward to see that the `flatter' the Higgs VEV (i.e. the smaller the value of $\alpha$), the smaller the last two terms of (\ref{P0eval}) will be and the smaller the contribution to the $T$ parameter will be\footnote{This should not be mistaken for a proof since we have not demonstrated that the $\alpha$ dependence of the last two terms dominates over the first two terms. It is probably possible to contrive a space for which this is not true. However, one can see that it will hold for all spaces in which $P_{W,Z}^{(0)}$ is growing towards the IR and $\partial_5 P_{W.Z}^{(0)}\propto \alpha^{n}$ with $n\geqslant0$.}. This should not be a particularly surprising result. One gets tree level corrections to the EW observables because the Higgs mixes the SM gauge fields (the zero modes) with the KK gauge fields that are peaked in the IR. The less the Higgs is peaked towards the IR then the weaker this mixing will be. Alternatively, in the language of a possible dual theory, the less composite the Higgs, the less it will mix the composite states with the elementary states. 

\begin{figure}[ht!]
\begin{center}
\includegraphics[width=0.8\textwidth]{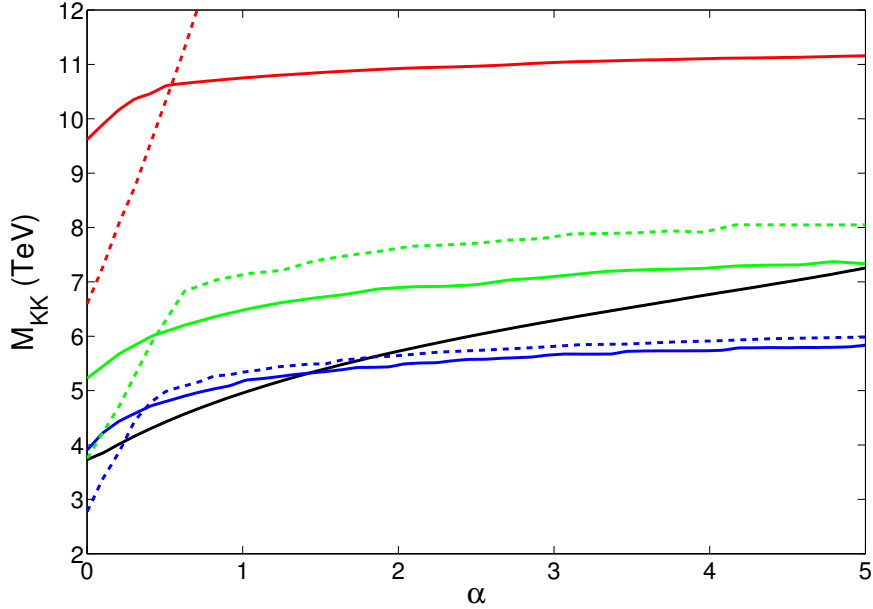}
\caption{The constraints on $M_{\rm{KK}}$ (not the mass of the first KK gauge boson) from EW precision tests for the RS model (black) as well as the modified metrics with $v=10$ (blue), $v=5$ (green) and $v=3$ (red). We have used $k\Delta=1$ (solid line), $k\Delta=0.5$ (dash dash), $m_H=125$ GeV, $M_{\rm{UV}}=k$ and fixed $\Omega=10^{15}$. At each iteration we have fit to $\hat{\alpha}$, $\hat{M}_Z$ and $\hat{G}_f$ and compared with the latest $S-T$ ellipse (\ref{STellipse}) at 95\% confidence level. For $k\Delta=0.5$, such an analysis is plagued by small numerical inaccuracies which results in not completely smooth curves.}
\label{fig:EWconstraints}
\end{center}
\end{figure}

The contribution to the first two terms of (\ref{P0eval}) are quite sensitive to the geometry of the space. It was found in \cite{Cabrer:2011vu, Cabrer:2011fb, Carmona:2011ib}, that the contribution to these two terms could be significantly reduced in the modified spaces (\ref{CGQMetric}) with $v\lesssim 1$ and $k\Delta\lesssim 1$. However, as discussed in section \ref{sect:HiggsVEV}, this region of parameter space typically results in the Higgs VEV growing more than exponentially which would increase the contribution from the last two terms of (\ref{P0eval}). So a minimum in the EW constraints can be found by introducing a moderate tuning such that the Higgs VEV still grows exponentially and $v\lesssim 1$. 

Before we can compute the size of these constraints, we must first perform a field redefinition in order to absorb the non-oblique corrections to the gauge-fermion couplings,
\begin{displaymath}
\tilde{W}_\mu\rightarrow\frac{g_4}{g\int dr \,ba^{-1}f_0^{(\psi)}G^{(W)}f_0^{(\psi)}}\tilde{W}_\mu,
\end{displaymath}
and likewise for the $Z$ field. Here $g_4$ is the 4D effective coupling found by fitting to (\ref{EWobserv}). This is only possible if one assumes universal fermion couplings and hence we again assume a universal fermion position with $c_L=-c_R=0.7$. In practice this is not a good approximation for the heavy quarks. Although the modified metrics, with more realistic fermion positions, have been studied in \cite{Carmona:2011ib} and it was still found that certain regions of the parameter space still have significantly reduced constraints. 

The constraints on the KK scale can then be calculated by comparison with a fit to the $S$ and $T$ parameters (assuming a vanishing $U$ parameter) \cite{Baak:2011ze},
\begin{equation}
\label{STellipse}
S=0.07\pm0.09\hspace{1cm} T=0.10\pm0.08\hspace{1cm} \rho_{\rm{correlation}}=+0.88.
\end{equation}        
These constraints have been plotted in figure \ref{fig:EWconstraints}. Note these constraints are for the KK scale and not for mass of the first KK gauge field. The masses of the first KK photon (or gluon) are given in table \ref{tab:KK masses}. Here we find that relatively small shifts in the geometry can result in significant reductions in the mass eigenvalues relative to the defined KK scale. It is believed that this is partly responsible for the reduction in the EW constraints found in \cite{Cabrer:2011vu, Cabrer:2011fb, Carmona:2011ib}. 

\section{The Pseudo-Scalars.}\label{sect:PseudoScal}
Having demonstrated that EW constraints will generically be at their minimum for smaller values of $\alpha$, we shall now move on to look at how models with a bulk Higgs can be potentially falsified. Before spontaneous symmetry breaking the model considered here contains four 5D massless gauge fields ($4\times 3$ transverse degrees of freedom) and the Higgs, a complex doublet ($4$ scalar degrees of freedom). After compactification and the breaking of EW symmetry the model contains four 4D massive gauge fields ($4 \times 2$ transverse degrees of freedom, plus $4$ longitudinal degrees of freedom) and a Higgs particle ($1$ scalar degree of freedom).  Hence the model must necessarily also include an additional three scalar degrees of freedom. In this section we shall demonstrate that, for $\alpha\sim\mathcal{O}(1)$ or less, such pseudo-scalars will gain masses at the KK scale and hence the observation of a $Z^{\prime}$ or $W^{\prime}$ at the LHC should be associated with the existence of a corresponding charged and neutral scalar. 

These pseudo-scalars have been previously investigated for warped spaces in \cite{Cabrer:2011fb, Falkowski:2008fz} and are well known in models with universal extra dimensions. As found in the appendix, these pseudo-scalars arise as a mixture of the $A_5$ component of the gauge field and the Higgs components, $\pi_i$, see (\ref{ phiZDef}) and (\ref{PhiWdef}). Note the larger the value of $\alpha$ the more these scalars will be dominated by the Higgs components $\pi_i$ and hence more of the $A_5$ component can be `gauged away'.

\begin{table}[t!]
  \centering 
  \begin{tabular}{|c||c|c|c|c|}
\hline
% after \\ : \hline or \cline{col1-col2} \cline{col3-col4} ...
   $k\Delta$& RS & $v=10$&$v=5$&$v=3$ \\
   \hline
   \hline
   0.5& 2.45  & 2.23 & 1.68 & 0.87 \\
   1&  2.45  & 2.27 & 1.82 & 1.07 \\
   1.5& 2.45 & 2.30 & 1.89 & 1.18 \\ 
\hline
\end{tabular}
  \caption{The mass of the first gauge boson relative to the KK scale, $m_1^{(A_\mu)}/M_{\rm{KK}}$.}\label{tab:KK masses}
\end{table}

\subsection{Consistent Boundary Conditions.} 
If we begin by considering a gauge field in the 5D space (\ref{genericMetric }) then variation of the action yields the boundary term
\begin{displaymath}
\left [\frac{1}{2}\delta A_\mu a^2b^{-1}\eta^{\mu\nu}F_{\nu 5}\right ]_{r=r_{\rm{IR}},r_{\rm{UV}}}=0.
\end{displaymath}
In order for such a boundary term to vanish one must impose either NBC's on the $A_\mu$ components and DBC's on the $A_5$ components or vice versa. If one is considering just an interval then one can, in theory, impose non-trivial boundary conditions. However if one is compactifying the space over a $S^1/Z_2$ orbifold, such non-trivial boundary conditions are forbidden by gauge invariance across the fixed points\footnote{To be explicit, the gauge field transforms over the orbifold as $A_\mu(r)\rightarrow A_\mu(-r)=PA_\mu(r) P^{\dag}$ and $A_5(r)\rightarrow A_5(-r)=-PA_5(r) P^{\dag}$. Where $P$ is a unitary matrix and the relative minus sign, which preserves gauge invariance under $A_M\rightarrow A_M-\frac{1}{g}\partial_M\Theta$, will ensure opposite BC's for $A_\mu$ and $A_5$.}. Hence if one imposes NBC's on the $A_\mu$ field, in order to gain a zero mode that is associated with the SM field, then one must impose DBC's on the $A_5$ component.     

 As for the Higgs particle, the consistent boundary conditions for the Higgs components are either DBC's, $\pi_i\big|_{r=r_{\rm{IR}},r_{\rm{UV}}}=0$, or non DBC's
 \begin{equation}
\label{piBCs}
\left [b^{-1}\partial_5\pi_i-M_{\rm{IR}}\pi_i+\lambda_{\rm{IR}}h^2\pi_i\right ]_{r=r_{\rm{IR}}}=0 \hspace{0.8cm}\mbox{and}\hspace{0.8cm}\left [b^{-1}\partial_5\pi_i-M_{\rm{UV}}\pi_i\right ]_{r=r_{\rm{UV}}}=0.
\end{equation}
However if we impose DBC's on the $A_5$ components then, in the unitary gauge, (\ref{Z5mix}) and (\ref{W5mix}) imply that one must impose DBC's on the pseudo-scalars, $\phi_{W,Z}\big|_{r=r_{\rm{IR}},r_{\rm{UV}}}=0$. To see that this is consistent we substitute (\ref{pi3mix}) and (\ref{piWMix}), 
\begin{equation}
\label{piExpans}
\pi_i=\sum_{n}\left (-\frac{M_{W,Z}a^{2}b^{-2}\partial_5\left (f_n^{(\phi_{W,Z})}\phi_{W,Z}^{(n)}\right )}{m_n^{(\phi_{W,Z})\,2}}\pm\frac{M_{W,Z}^{-1}a^{-2}b^{-1}\partial_5\left (a^{4}b^{-1}M_{W,Z}^2\right )}{m_n^{(\phi_{W,Z})\,2}}f_n^{(\phi_{W,Z})}\phi_{W,Z}^{(n)}\right )
\end{equation}
where the $\pm$ refers to $\pi_{1,3}$ and $\pi_2$ respectively, into the boundary conditions of $\pi_i$ and note that the second term in (\ref{piExpans}) will vanish by the DBC's. While if we impose DBC's on the Higgs VEV, $h$, then the first term will always vanish. Alternatively if we impose non DBC's (\ref{HiggVEVBCs}) then we must impose impose non DBC's on the Higgs components (\ref{piBCs}).   

The upshot of all this is that if NBC's are imposed on the $A_\mu$ field then the pseudo-scalars must have DBC's, regardless of the boundary conditions of the Higgs. This should be intuitive if one considers the zero mode of the $A_\mu$ field which will be a massless 4D gauge field and hence will have 2 transverse degrees of freedom after compactification. By the same arguments, that we began this section with, then clearly the degrees of freedom would not match if the physical pseudo-scalars also gained a zero mode. 

\subsection{The Pseudo-Scalar Masses.}
\begin{figure}
\begin{center}
\includegraphics[width=0.8\textwidth]{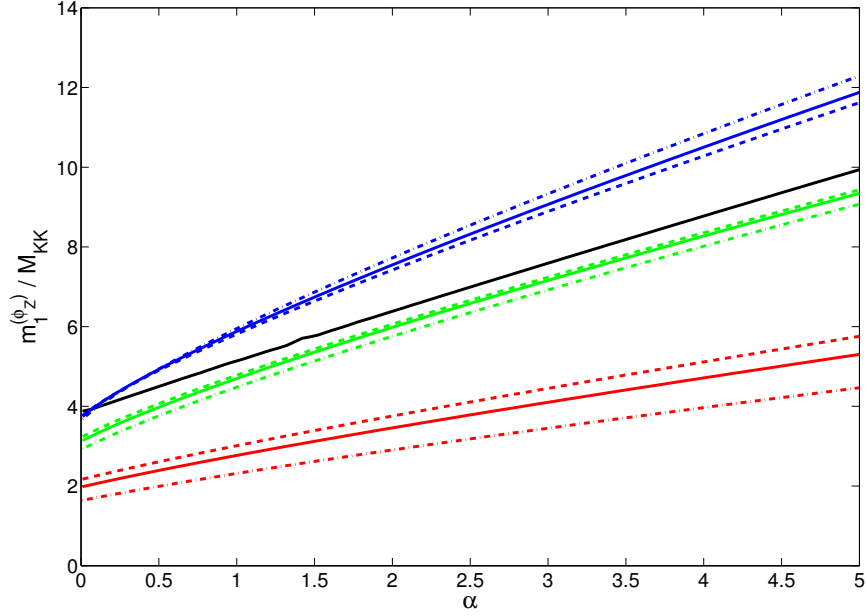}
\caption{The masses of the first Z pseudo scalar for the RS model (black), as well as the modified metrics with $v=10$ (blue), $v=5$ (green) and $v=3$ (red) and $k\Delta=0.5$ (dash-dot line), $k\Delta=1$ (solid line) and $k\Delta=1.5$ (dash-dash line). }
\label{fig:ScalarMasses}
\end{center}
\end{figure}

For gauge fields propagating in (\ref{MetricUsed}), the pseudo-scalar masses and profiles are found by solving (\ref{pseudoEOMZ}) and (\ref{pseudoEOMW}),
\begin{eqnarray}
\label{phiEOMusedMet}
\partial_5^2f_n^{(\phi_{W,Z})}+\left (-6A^{\prime}+\frac{2M_{W,Z}^{\prime}}{M_{W,Z}}\right )\partial_5f_n^{(\phi_{W,Z})}\hspace{10cm}\nonumber\\
+\left (-4A^{\prime\prime}+8A^{\prime\,2}+2\frac{M_{W,Z}^{\prime\prime}}{M_{W,Z}}-2\frac{M_{W,Z}^{\prime\,2}}{M_{W,Z}^2}-4A^{\prime}\frac{M_{W,Z}^{\prime}}{M_{W,Z}}-M_{W,Z}^2+e^{2A}m^{(\phi_{W,Z})\,2}_n\right )f_n^{(\phi_{W,Z})}=0\label{phiEOMusedMet}
\end{eqnarray}
where ${}^{\prime}$ denotes $\partial_5$. Typically this equation cannot be solve analytically and hence we must work with numerics. The first pseudo-scalar masses have been plotted in figure \ref{fig:ScalarMasses}. For the RS model (\ref{AdSMetric}) and (\ref{RSHiggsVEV}), this gives
\begin{displaymath}
\frac{M_{W,Z}^{\prime}}{M_{W,Z}}=2k+\alpha k\frac{e^{\alpha kr}-Be^{-\alpha kr}}{e^{\alpha kr}+Be^{-\alpha kr}}\hspace{0.8cm}\mbox{and}\hspace{0.8cm}\frac{M_{W,Z}^{\prime\prime}}{M_{W,Z}}=(4+\alpha)k^2+4\alpha k^2\frac{e^{\alpha kr}-Be^{-\alpha kr}}{e^{\alpha kr}+Be^{-\alpha kr}}
\end{displaymath}
and so as $\alpha\rightarrow 0$ then (\ref{phiEOMusedMet}) will reduce to the same equations of motion of a gauge field of mass $M_{W,Z}$ (\ref{ZEOM}), i.e the mass eigenfunctions will be approximately root one Bessel functions. Also as either $\alpha\rightarrow \infty$ (i.e. a brane localised Higgs) then $\frac{M_{W,Z}^{\prime}}{M_{W,Z}}\rightarrow \infty$ and the pseudo-scalars will become infinitely heavy. Likewise if $g\rightarrow 0$ (i.e. the gauge fields become decoupled from the Higgs) then $\phi_{W,Z}\rightarrow 0$ and the field should be considered unphysical. It is also clear from figure \ref{fig:ScalarMasses}, as well as the above equations, that as $\alpha$ is increased the mass of the pseudo-scalars will always increase. Hence, at tree level, the pseudo-scalars will always be heavier than the first KK gauge field. This feature further adds to the ability of the LHC to rule out this model.

\subsection{The Pseudo-Scalar Couplings to Fermions.}
We can now move on to look at how the pseudo-scalars couple to fermions. Again in the interests of generality we shall work with the generic metric (\ref{genericMetric }). The Higgs sector and gauge sector couple to the SM fermions via the Yukawa couplings and the fermion kinetic terms
\begin{equation}
\label{ }
S=\int d^5x \bigg ( i\frac{b}{a}\bar{\Psi}\gamma^\mu D_\mu \Psi+i\bar{\Psi}\gamma^5 D_5 \Psi+i\frac{b}{a}\bar{X}\gamma^\mu D_\mu X+i\bar{X}\gamma^5 D_5 X -bY_D\bar{\Psi}\Phi X-bY_U\epsilon_{ab}\bar{\Psi}^a\Phi^{\dag b}X +h.c.\bigg ),
\end{equation}
where $\Psi\ni\{L,Q\}$ and $X\ni\{e,u,d\}$, as in section \ref{sect:FermMass}. The first and third $\gamma^\mu D_\mu$ terms are just the usual gauge fermion interactions. The second and fourth $\gamma^5D_5$ terms will give rise to pseudo-scalar currents interacting with the $A^a_5$ components, for example. With $\gamma^5=\left(\begin{array}{cc}i & 0 \\0 & -i\end{array}\right)$ then
\begin{displaymath}
i\bar{\Psi}\gamma^5 D_5 \Psi\supset i\bar{\Psi}\gamma^5(-ig\tau^aA^a_5)\Psi=-ig\bar{\psi}_L\tau^aA^a_5\psi_R+ig\bar{\psi}_R\tau^aA^a_5\psi_L,
\end{displaymath}
where $A_5^a$ are the fifth components of the SU$(2)$ gauge bosons, as in the appendix. If we remember that, after compactification over the orbifold, only the left handed components of $\Psi $ and the right handed components of $X$ will gain a zero mode then clearly the $A_5^a$ component will only couple to currents containing one SM fermion and one KK fermion or alternatively two KK fermions. Hence at tree level, the pseudo-scalars will only couple to SM fermion currents via the Yukawa couplings to the Higgs components, $\pi_i$. Bearing in mind that the larger the value of $\alpha$, the more the pseudo-scalars will be dominated by the $\pi_i$ components and hence we can see our first indication  that pseudo-scalars will have more significance for SM processes, for larger values of $\alpha$.

At next to leading order, the phenomenology becomes more complex and a proper investigation is beyond the scope of this paper. Although it is worth commenting that the Higgs particle will couple to the pseudo-scalars via the $|\Phi|^4$ term and so one would anticipate that the Higgs mass will be sensitive to the pseudo-scalar masses. Hence if $\alpha$ is very large and the pseudo-scalars are very heavy, then one risks reintroducing a little hierarchy problem. 

\begin{figure}[ht!]
    \begin{center}
        \subfigure[The effective coupling of the first $\phi_Z$ field to the SM fermions assuming $\tilde{Y}=1$.]{%
            \label{fig:Yeff}
            \includegraphics[width=0.7\textwidth]{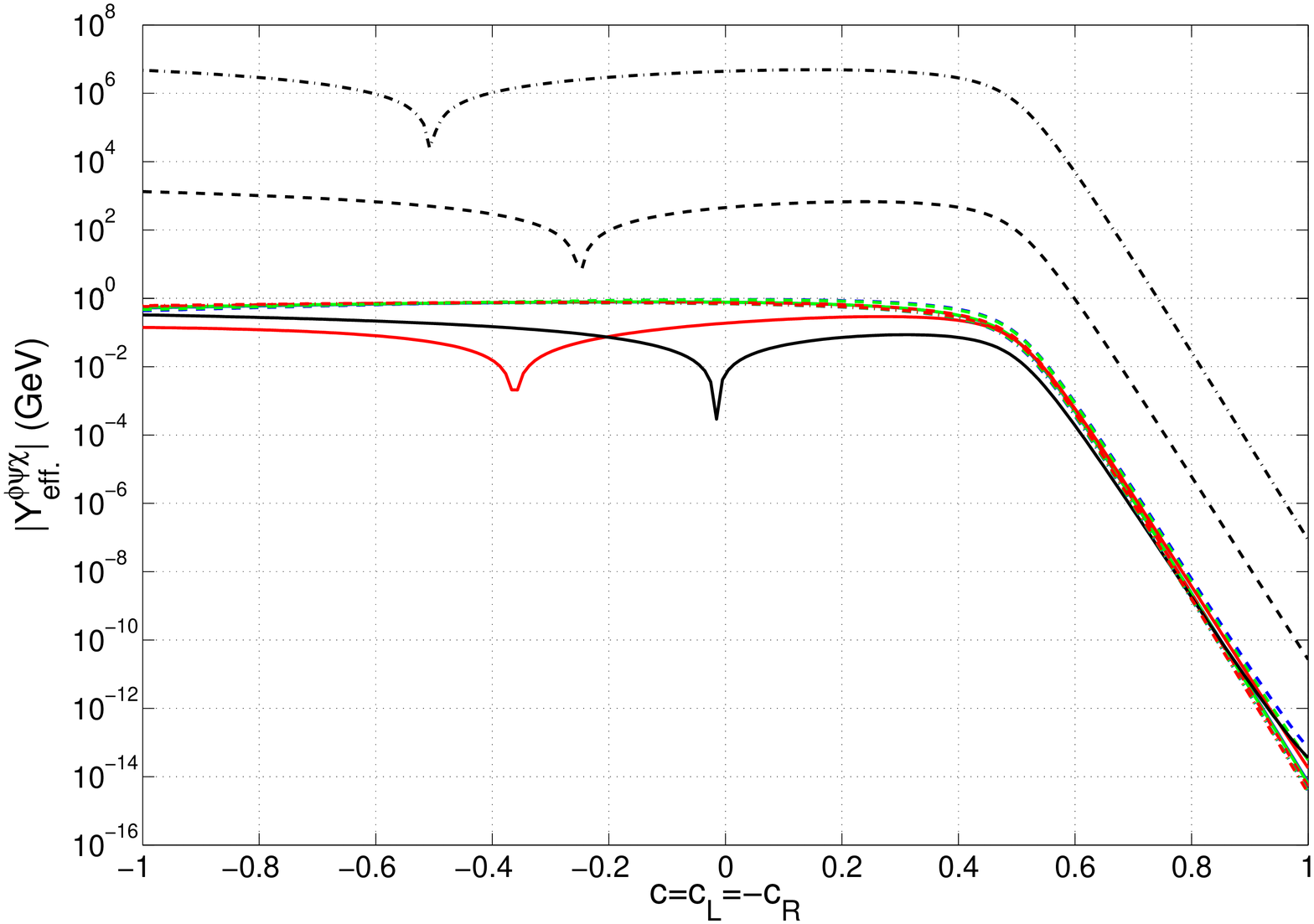}
        }\\
        \subfigure[The effective Yukawa coupling relative to the fermion mass. The less universal this value is, then the greater the pseudo-scalar mediated FCNC's will be.]{%
           \label{fig:RelativeYeff}
           \includegraphics[width=0.7\textwidth]{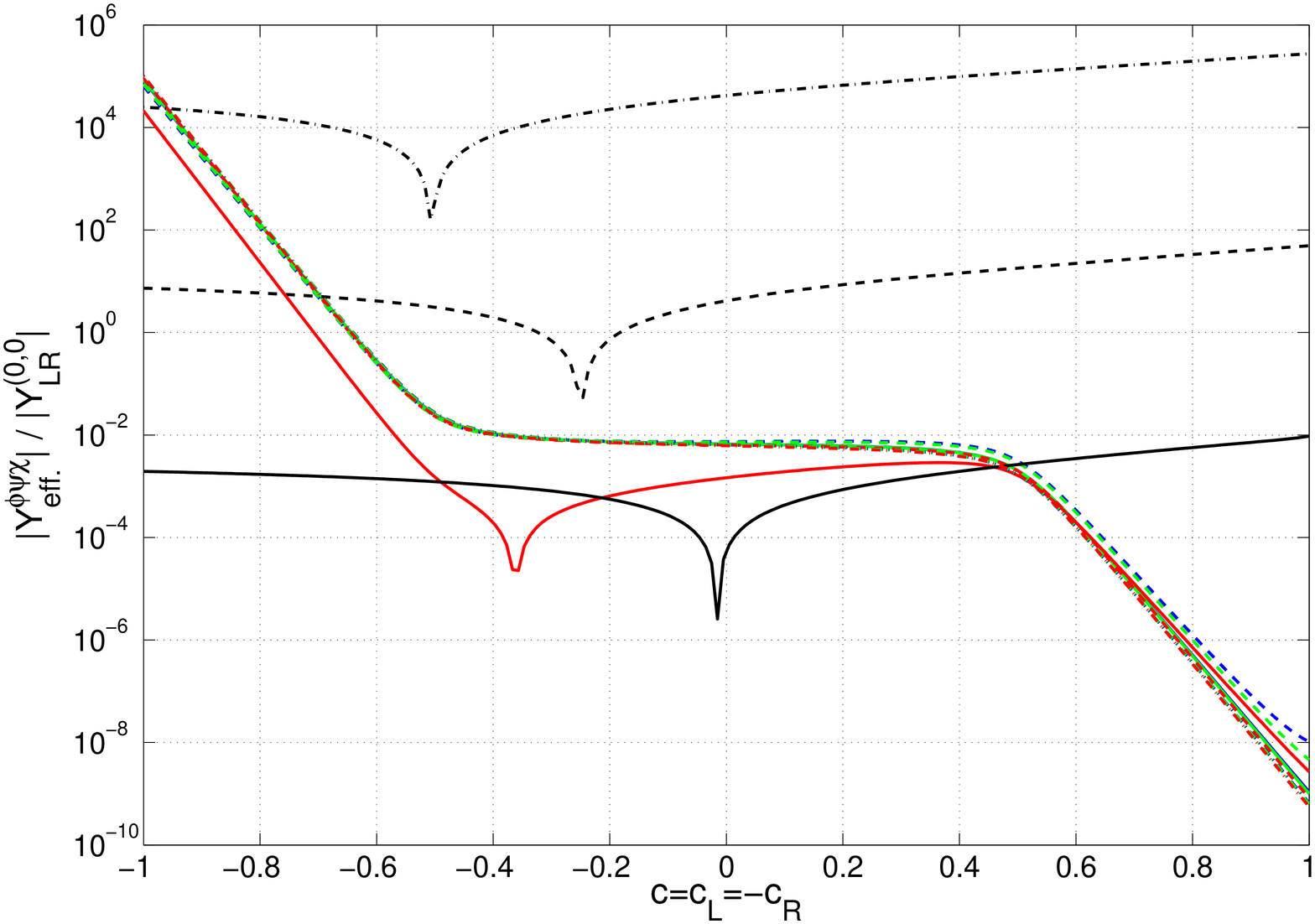}
        }
    \end{center}
    \caption{The couplings of the Z pseudo-scalar to SM fermions, with bulk mass terms $c$, for the RS model (black) and modified metrics with $v=10$ (blue), $v=5$ (green) and $v=3$ (red). Here we fix $k\Delta=1$ and consider $\alpha =0.01$ (solid line), $\alpha=0.5$ (dash-dash line) and $\alpha=1.01$ (dash-dot line). $\Omega=10^{15}$ and $M_{\rm{KK}}=2$ TeV.} \label{fig:ScalarCoupl}
\end{figure}

Moving onto look at the Yukawa couplings and looking just at the quarks,
\begin{eqnarray}
S=\int d^5x\; \frac{b}{\sqrt{2}}\bigg(-Y_D\;\bar{u}_{(\psi)}\pi_{+}d_{(\chi)}-Y_D\;\bar{d}_{(\psi)}(h+H+i\pi_3)d_{(\chi)}\hspace{2cm}\nonumber\\
-Y_U\;\bar{u}_{(\psi)}(h+H-i\pi_3)u_{(\chi)}-Y_U\;\bar{d}_{(\psi)}\pi_{-}u_{(\chi)} +h.c.\bigg )
\end{eqnarray} 
where $\pi_{\pm}=\frac{1}{\sqrt{2}}(\pi_1\pm i\pi_2)$, we have also split $Q=\left(\begin{array}{c}u_{(\psi)} \\d_{(\psi)}\end{array}\right)$ and the $(\psi,\chi)$ subscripts refer to whether the fermion originated from a SU$(2)$ doublet or a singlet. To find the effective coupling to the SM fermions, we can now plug in (\ref{pi3mix}), (\ref{piWMix}), carry out a partial integration and use the DBC's of $\phi_{W,Z}$ to get
\begin{equation}
\label{ }
\mathcal{L}_{\rm{eff.}}\subset -\frac{Y}{\sqrt{2}}\left (\int dr\; \frac{a^4b^{-1}M_{W,Z}^2f_n^{(\phi_{W,Z})}}{m_n^{(\phi_{W,Z})\,2}}\partial_5\left (M_{W,Z}^{-1}a^{-2}f_0^{(\psi_L)}f_0^{(\chi_R)}\right )\right )\;\phi_{W,Z}^{(n)}\psi_L^{(0)}\chi_R^{(0)}\;\equiv Y_{\rm{eff.}}^{(\phi_{W,Z}\psi\chi)}\phi_{W,Z}^{(n)}\psi_L^{(0)}\chi_R^{(0)}.
\end{equation}
For the spaces considered in this paper (\ref{MetricUsed}), this gives
\begin{equation}
\label{ }
Y_{\rm{eff.}}^{(\phi_W\psi\chi)}=-\frac{gYN_\psi N_\chi}{2\sqrt{2}\,m_n^{(\phi_{W})\,2}}\int dr\; \left (2A^{\prime}h-h^{\prime}-(c_L-c_R)kh\right )e^{-2A-(c_L-c_R)kr}f_n^{(\phi_W)},
\end{equation}
where $N_{\psi ,\chi}$ are the fermion normalisation constants. At first sight, one may be alarmed to see that this effective Yukawa coupling appears to be dependent on the gauge coupling, $g$. However this is not the case since the normalisation constant of $f_n^{(\phi_W)}$, obtained from (\ref{PhiOrthogRel}), will contain a factor of $g^{-1}$. Likewise the effective coupling will be independent of $N_h$ at tree level. This is an interesting coupling for a number of reasons. Firstly it gives a good example of a scenario in which relatively small changes in the geometry can result in significant changes in the phenomenology. In the RS model, the `$e^{2kr}$ factor' in the $h^{\prime}$ term will exactly cancel with the $2A^{\prime}h$ term and the effective coupling will be given by  
\begin{eqnarray}
Y_{\rm{eff.}}^{(\phi_W\psi\chi)}=-\frac{YgN_hk^2}{2\sqrt{2}}\sqrt{\frac{(1-2c_L)(1+2c_R)}{(\Omega^{1-2c_L}-1)(\Omega^{1+2c_R}-1)}}\frac{1}{m_n^{(\phi_W)\,2}} \hspace{6cm}\nonumber\\
\hspace{1cm}\times\int dr\; f_n^{(\phi_W)}\left (e^{(-c_L+c_R)kr}\left ((-c_L+c_R)(e^{\alpha kr}+Be^{-\alpha kr})-\alpha(e^{\alpha kr}-Be^{-\alpha kr})\right )\right ).\hspace{-0.5cm}
\end{eqnarray} 
This coupling is very sensitive to the value of $\alpha$, as can be seen in figure \ref{fig:ScalarCoupl}.  For large values of $\alpha$ the pseudo-scalars would become strongly coupled and one would loose perturbative control of the theory. On the other hand, this cancellation does not occur in spaces with a modified metric. The resulting effective coupling has additional terms which result in the coupling being significantly less $\alpha$ dependent.    
  
The second interesting feature of this coupling is that there exists special fermion positions (when $2A^{\prime}h-h^{\prime}-(c_L-c_R)kh\approx0$) for which the pseudo-scalars tree level coupling, to the SM fermions, is significantly reduced. This is analogous to the case of the KK gauge fields coupling to SM fermions, when $c_L=-c_R=0.5$. These fermion positions typically occur when the fermions are heavily peaked towards the IR and for many of the modified metrics considered, this occurs for bulk mass parameters not plotted in figure \ref{fig:ScalarCoupl}. 

Clearly there is significant room for further work concerning the phenomenological implications of these pseudo-scalars. However by going further in this direction we risk digressing too far from the central focus of this paper, that of investigating the fermion mass hierarchy and looking for possible motivations for considering a small value of $\alpha$ to be more plausible than a large value. Nevertheless, since we have suggested that the observation, or lack of observation, of such scalars could play an important role in excluding this model it is worth making some comments concerning their phenomenology. In addition to the Yukawa couplings, the pseudo scalars will also couple to the $\gamma$, $W$ and $Z$ gauge fields, the KK gravitons, the radion and the Higgs itself. Also, due to their large masses, one would anticipate that they could decay via a KK particle. Hence any study of the production and decay of such particles would be quite involved and should really be conducted as a separate piece of work.   
 
Returning to the focus of this paper, after one has diagonalised the fermion mass matrix (\ref{FermionMassMatrix}), this model would give rise to the possibility of tree level pseudo-scalar mediated FCNC's. The constraints from these FCNC's have the potential to force the KK scale to a level at which it becomes questionable whether one has resolved the hierarchy problem. The extent to which such FCNC's would be suppressed is determined by the size of the effective Yukawa couplings  as well as the size of its misalignment with the fermion masses or equivalently its universality with respect to fermion position. This has been plotted in figure \ref{fig:ScalarCoupl}. It is found that, when the fermions are peaked towards the UV ($c>0.5$), although this coupling is significantly misaligned and non-universal, it is also very small and hence such FCNC's would be heavily suppressed. However, for the RS model, as $\alpha$ is increased the size of this coupling increases and it rapidly becomes potentially problematic. Surprisingly, this effect is not as severe in spaces with a modified metric, for reasons discussed above.     

\section{Implications for Flavour Physics.}\label{sect:flavour}
Of course, this model would not just receive constraints from pseudo-scalar mediated FCNC's. It is well known that models with warped extra dimensions, that seek to explain the fermion mass hierarchy, suffer from severe constraints from flavour physics (see for example \cite{Huber:2003tu, Agashe:2004cp, Csaki:2008zd, Blanke:2008zb, Casagrande:2008hr, Bauer:2009cf}). These constraints fall into three broad categories:
\begin{itemize}
 \item Those that are dominated by the modification of a SM coupling. These include, for example, corrections to the $Z\bar{b}b$ vertex as well as rare leptonic decays such as $\mu\rightarrow ee\bar{e}$.
  \item Those that are dominated by the tree level exchange of a KK particle. For example, a particularly stringent constraint arises from $\epsilon_K$ that receives large contributions from the exchange of a KK gluon. 
  \item Those that arise at next to leading order via additional penguin diagrams. In particular, constraints from $\mu\rightarrow e\gamma$ \cite{Agashe:2006iy, Csaki:2010aj} and $b\rightarrow s\gamma$ \cite{Blanke:2012tv} transitions. 
\end{itemize}

\begin{figure}[ht!]
\begin{center}
\includegraphics[width=0.85\textwidth]{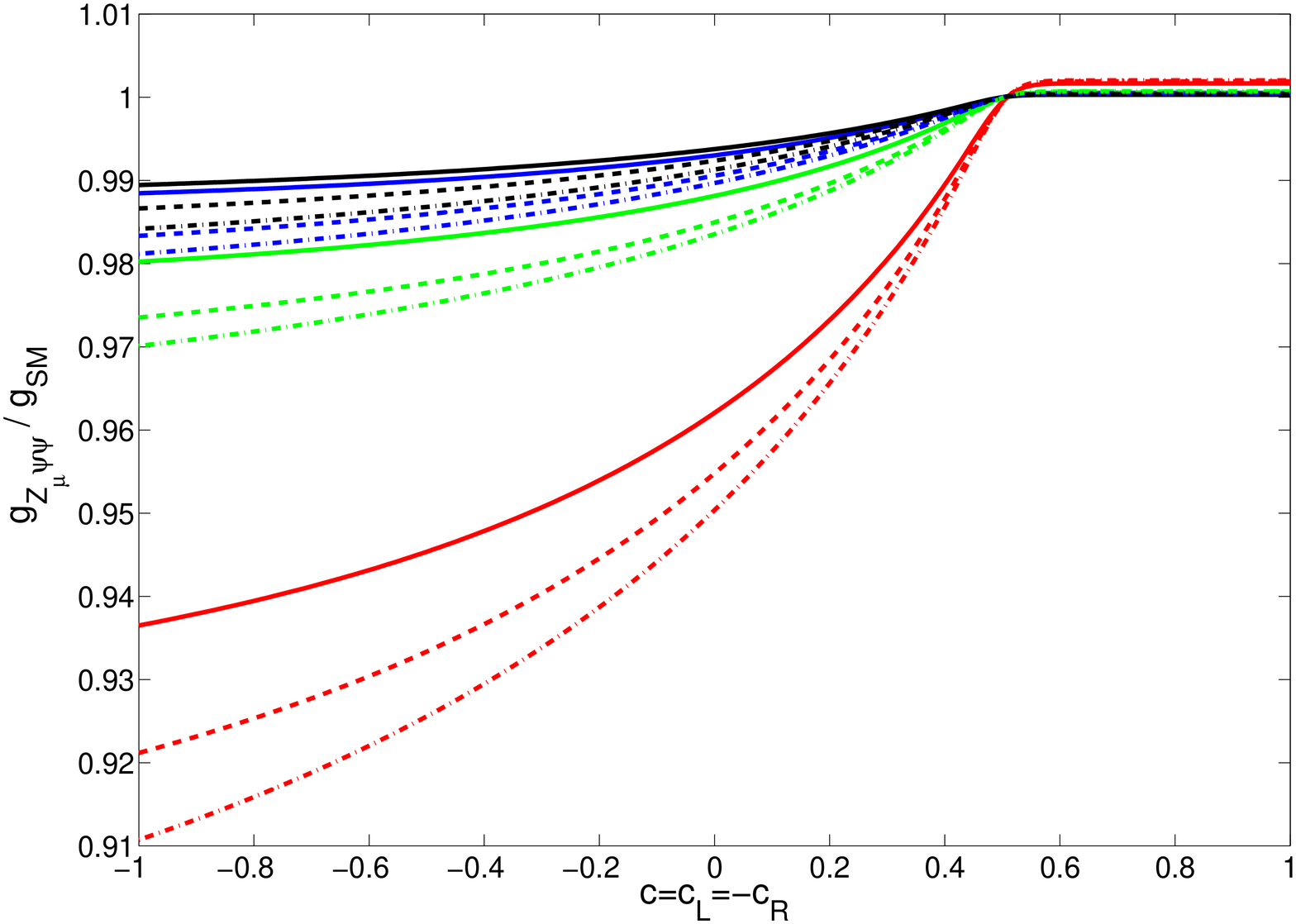}
\caption{Coupling of the Z zero mode to fermions at positions $c=c_L=-c_R$ for the RS model (black) and the IR modified spaces with $k\Delta=1$ and $v=10$ (blue), $v=5$ (green), $v=3$ (red). We use the Higgs exponent of $\alpha =0.01$ (solid lines), $\alpha=1.01$ (dash-dash line) and $\alpha=5$ (dash-dot line). In the interests of making a fair comparison we take a KK scale such that the mass of the first KK gauge field is the same, i.e. for RS $M_{\rm{KK}}=2$ TeV, for $v=10$ $M_{\rm{KK}}=2.15$ TeV, for $v=5$ $M_{\rm{KK}}=2.70$ TeV and for $v=3$ $M_{\rm{KK}}=4.59$ TeV. Also $\Omega=10^{15}$. }
\label{fig:Zcoupl}
\end{center}
\end{figure}

While it certainly lies beyond the scope of this paper to conduct a full and thorough investigation of flavour in models with a bulk Higgs, we are in a position to see how the first two categories of constraints would be effected. It is more difficult to estimate the effect on the next to leading order processes due to the number of conflicting factors contributing to these constraints. In particular there would be modifications to both the KK masses and the couplings plus additional diagrams coming from the pseudo-scalars and the KK Higgs bosons. 

\subsection{Modifications to the SM Fermion Couplings.}
Of particular importance, to many constraints from flavour physics, is the modification to the $Z$ fermion coupling, which has been plotted in figure \ref{fig:Zcoupl}. This coupling not only adds a direct constraint on the KK scale, via rare lepton decays and the partial decay width of $Z\rightarrow\bar{b}{b}$, it also constrains the extents to which a split fermion scenario is allowed. Such a modification occurs because, after spontaneous symmetry breaking, the $Z$ zero mode is not flat but modified in the IR. This gives rise to non-universal couplings to fermions in different locations. As one would expect, the flatter the Higgs VEV the smaller this modification will be. Hence the smaller the value of $\alpha$ the smaller these constraints will be. It is this effect which is partly responsible for the reduction in the constraints from rare lepton decays seen in \cite{Archer:2011bk, Atkins:2010cc}. Although it is also found that modifications to the geometry typically give an enhancement of the deformation of the $Z$ zero mode, relative to AdS${}_5$. 

The fermion zero mode profiles will also receive corrections arising from the mixings between the SU$(2)$ singlets and doublets, see (\ref{fermCorrections}). These corrections can also give a sizeable correction to the          
$Z\rightarrow\bar{b}{b}$ vertex \cite{Casagrande:2008hr}. The size of these corrections are determined by the size of $Y_{LR}^{(0,n)}$ for $n\geqslant 1$. Whether this correction increases or decreases, as one increases $\alpha$, is not entirely clear since it is dependent on whether the fermion zero mode profiles are peaked towards the UV brane or towards the IR brane. Hence it is sensitive to the preferred fermion positions (see the following section). It is also sensitive to how quickly the $Y_{LR}^{(0,n)}$ decreases as one increases $n$. It is straight forward to check that $Y_{LR}^{(0,n)}$ will drop away more quickly for smaller values of $\alpha$. In other words, as one shifts the Higgs VEV away from the IR brane one reduces the mixing between fermion zero modes and the higher KK fermion modes.   

Another source of flavour violation, in these models, would be the Higgs particle couplings to the SM fermions. Since the Higgs particle profile (\ref{HiggsPartProfile}) is not the same as the Higgs VEV (\ref{RSHiggsVEV}), after diagonalising (\ref{FermionMassMatrix}), the resulting Higgs particle effective Yukawa couplings will not be flavour diagonal \cite{Azatov:2009na}. The size of this effect will be determined by the difference between the Higgs VEV and the Higgs profile which, after expanding (\ref{HiggsPartProfile}), is found to be $\sim\mathcal{O}\left (\left (m_0^{(H)}e^{kr}/k\right )^2 \frac{1}{2\alpha}\right )$. Hence this is the one counter example considered in which reducing $\alpha$ would increase the size of the correction. However in practice it is found, for all spaces considered, that when $m_0^{(H)}\ll M_{\rm{KK}}$ this effect is quite small.  

\subsection{FCNC's from the Exchange of a KK Particle.}
\begin{figure}[t!]
    \begin{center}
        \subfigure[Relative coupling of a KK gluon to SM fermions.]{%
            \label{fig:GluonCoupling}
            \includegraphics[width=0.48\textwidth]{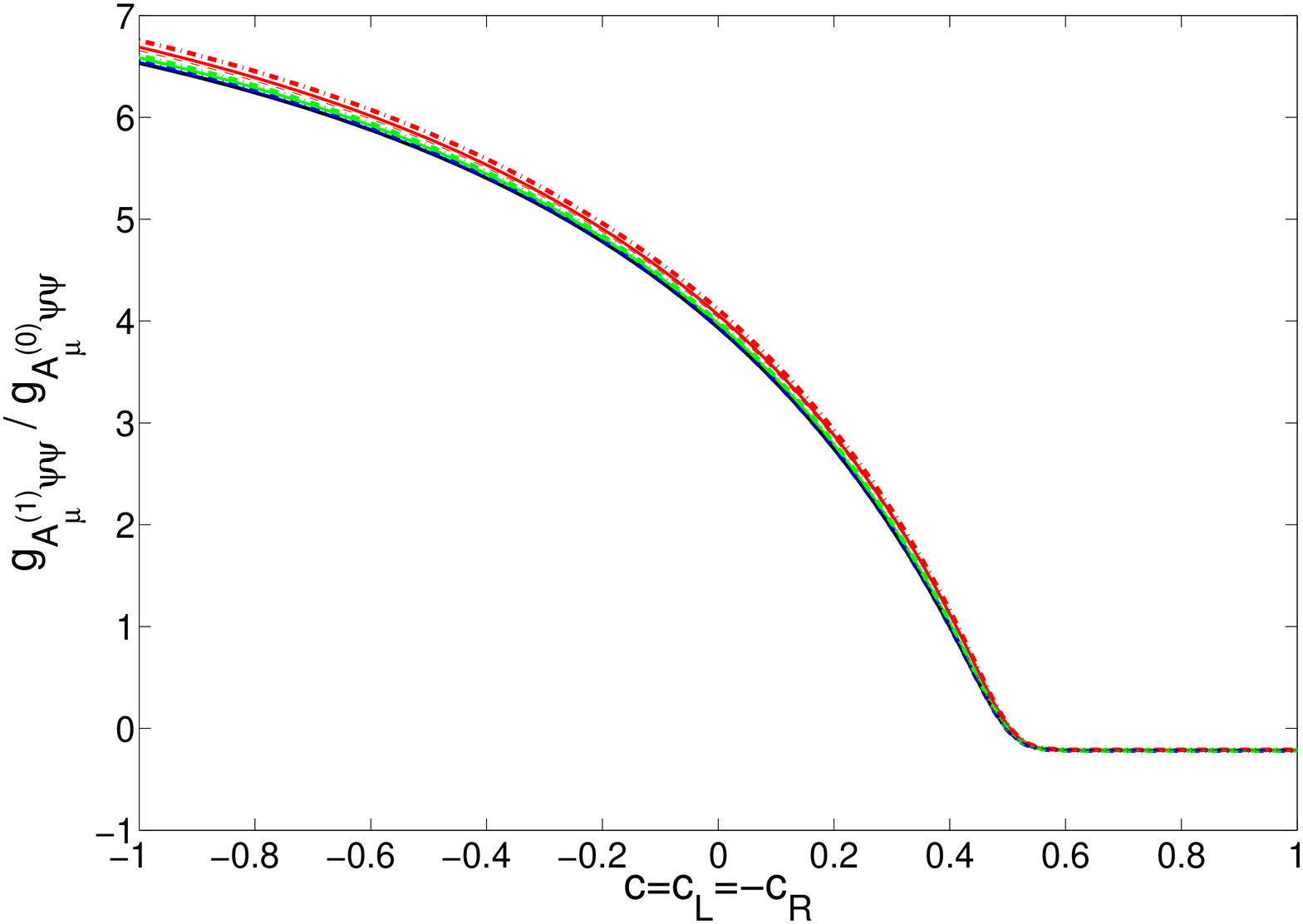}
        }
        \subfigure[Universality of  the coupling to fermions peaked towards UV.]{%
           \label{fig:GluonUniversality}
           \includegraphics[width=0.48\textwidth]{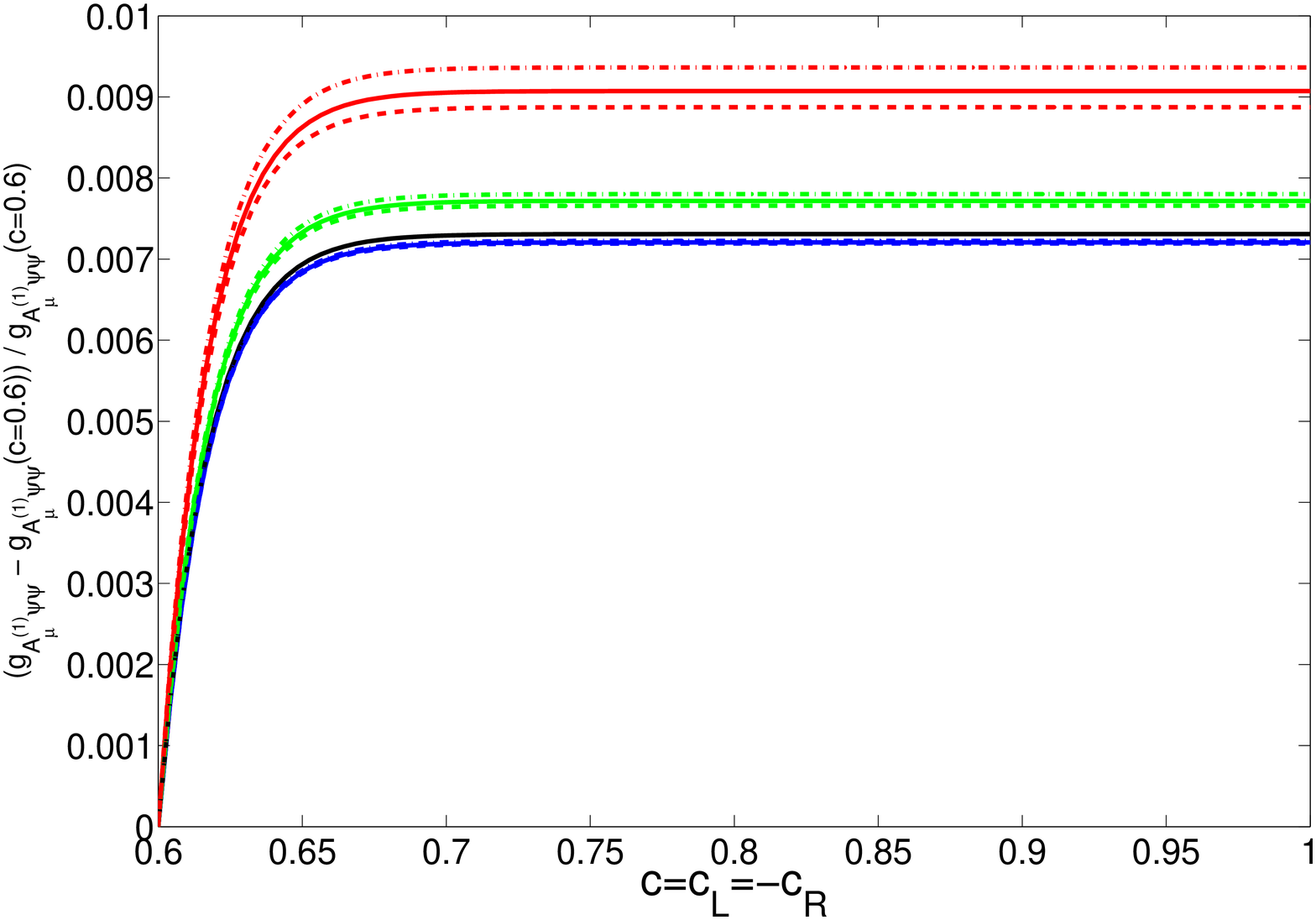}
        }
    \end{center}
    \caption{Plotted is the relative coupling of a KK gluon for the RS model (black) and the modified metrics with $v=10$ (blue), $v=5$ (green) and $v=3$ (red). As well as with $k\Delta=1.5$ (dash-dash line), $k\Delta=1$ (solid line) and $k\Delta=0.5$ (dash-dot line). These couplings are independent of $\alpha$ at tree level. The relative difference in the couplings between fermions peaked towards the UV and a fermion with a bulk mass parameter of $c=0.6$ has been plotted on the right. $\Omega=10^{15}$ and $M_{\rm{KK}}=2$ TeV. } \label{gluCoupl}
\end{figure}

As already mentioned, models with warped extra dimensions and bulk fermions suffer from stringent constraints from FCNC's arising at tree level via the exchange of a KK particle. In particular from $K^0-\bar{K}^0$ mixing and the observable $\epsilon_K$. Such FCNC's occur at tree level because the KK gluon has a non-universal coupling to fermions with different bulk mass parameters. Models with warped extra dimensions and a large warp factor have a natural suppression of such FCNC's referred to as the RS-GIM mechanism (see for example \cite{Huber:2003tu}). In particular fermions peaked towards the UV ($c>0.5$) have approximately universal couplings to the KK gluon, see figure \ref{gluCoupl}. The problem really arises because typically it is found that the heavier quarks have to sit quite far towards the IR and hence have non-universal couplings.

The effect on this constraint, of allowing the Higgs to propagate in the bulk, has been studied for the RS model \cite{Agashe:2008uz}, spaces with a modified metric \cite{Cabrer:2011qb} and spaces with a soft wall \cite{Archer:2011bk}. All three papers found a sizeable reduction in the constraint relative to that of the RS model with a brane localised Higgs. It is easily checked that the couplings of a KK gluon propagating, in the modified metrics, are not significantly modified relative to that of the RS model, see figure \ref{gluCoupl}. This coupling is also found to be independent of $\alpha$. Hence it is suspected that the primary reason for this reduction in the $\epsilon_K$ constraint is due to a shift in the preferred fermion positions.  In particular the heavy quarks can sit further towards the UV, where the coupling is more universal. When referring to the preferred fermion positions we are referring to the bulk mass parameters, $c_{L,R}$, that give the correct masses and mixing angles assuming anarchic order one 5D Yukawa couplings, $\tilde{Y}$.

\begin{figure}[ht!]
    \begin{center}
        \subfigure[$c_L^3$]{%
            \label{fig:cL3}
            \includegraphics[width=0.45\textwidth]{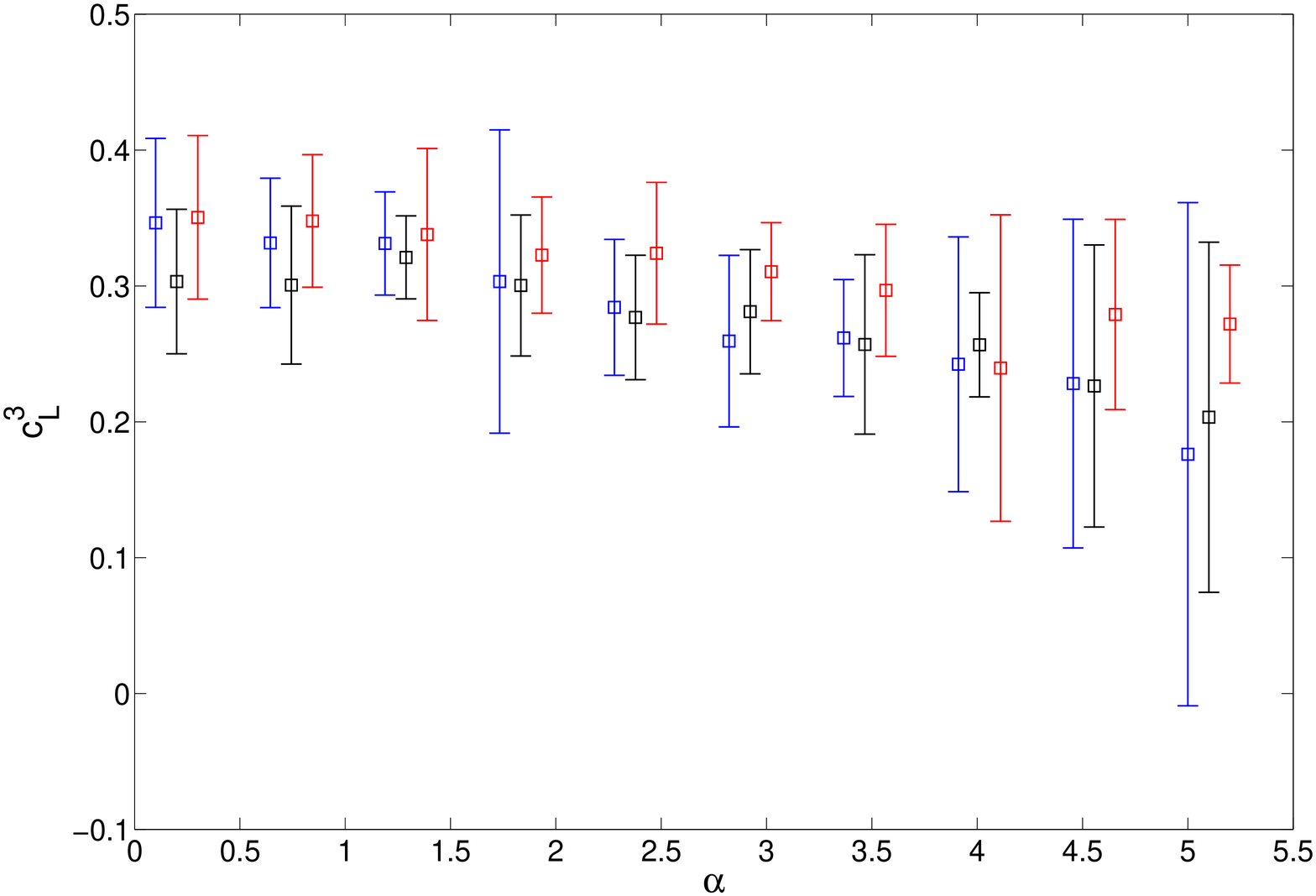}
        }
        \subfigure[$c_L^2$]{%
           \label{fig:cL2}
           \includegraphics[width=0.45\textwidth]{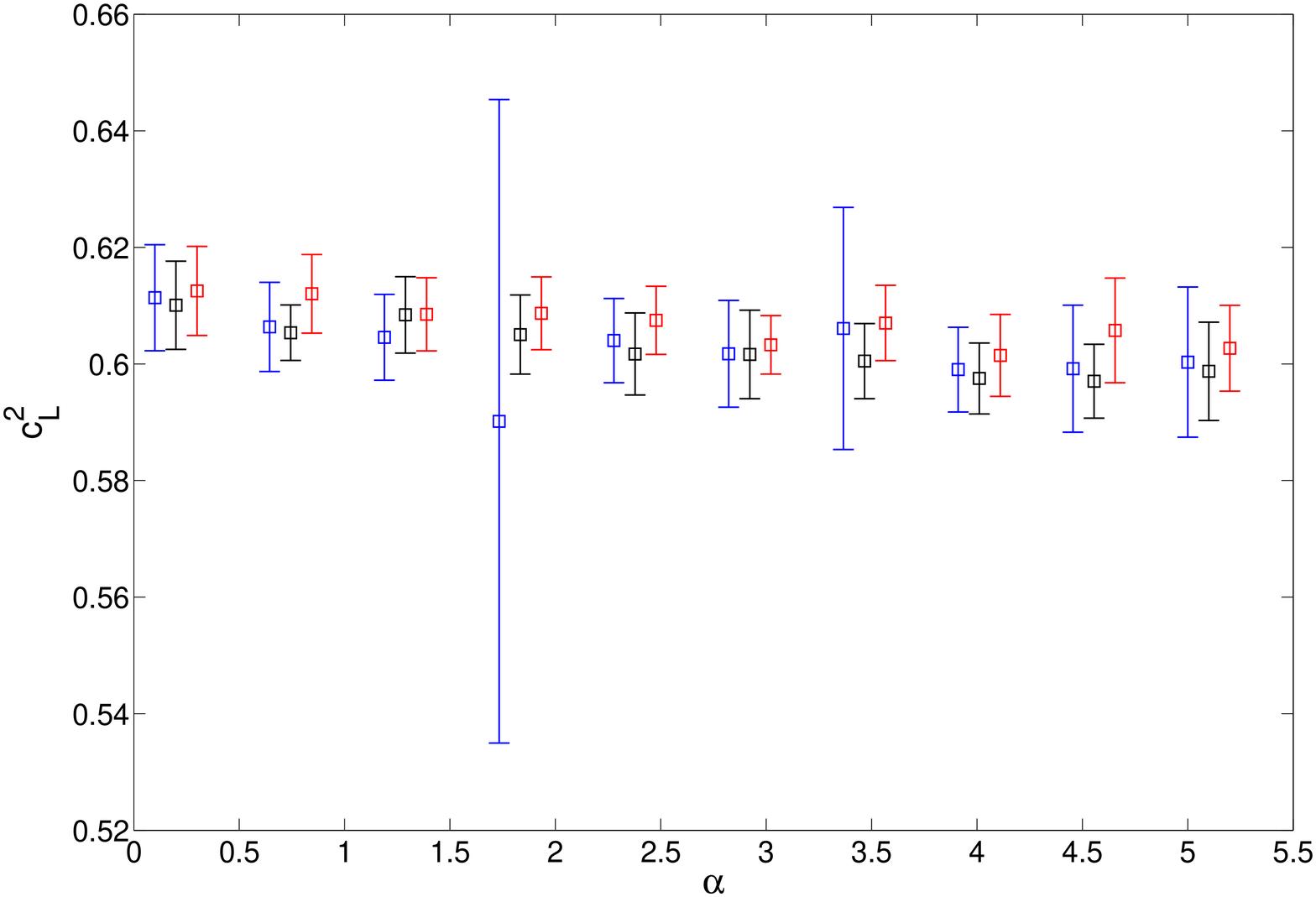}
        }\\
    \subfigure[$c_R^{u\, 3}$]{%
            \label{fig:uR3}
            \includegraphics[width=0.45\textwidth]{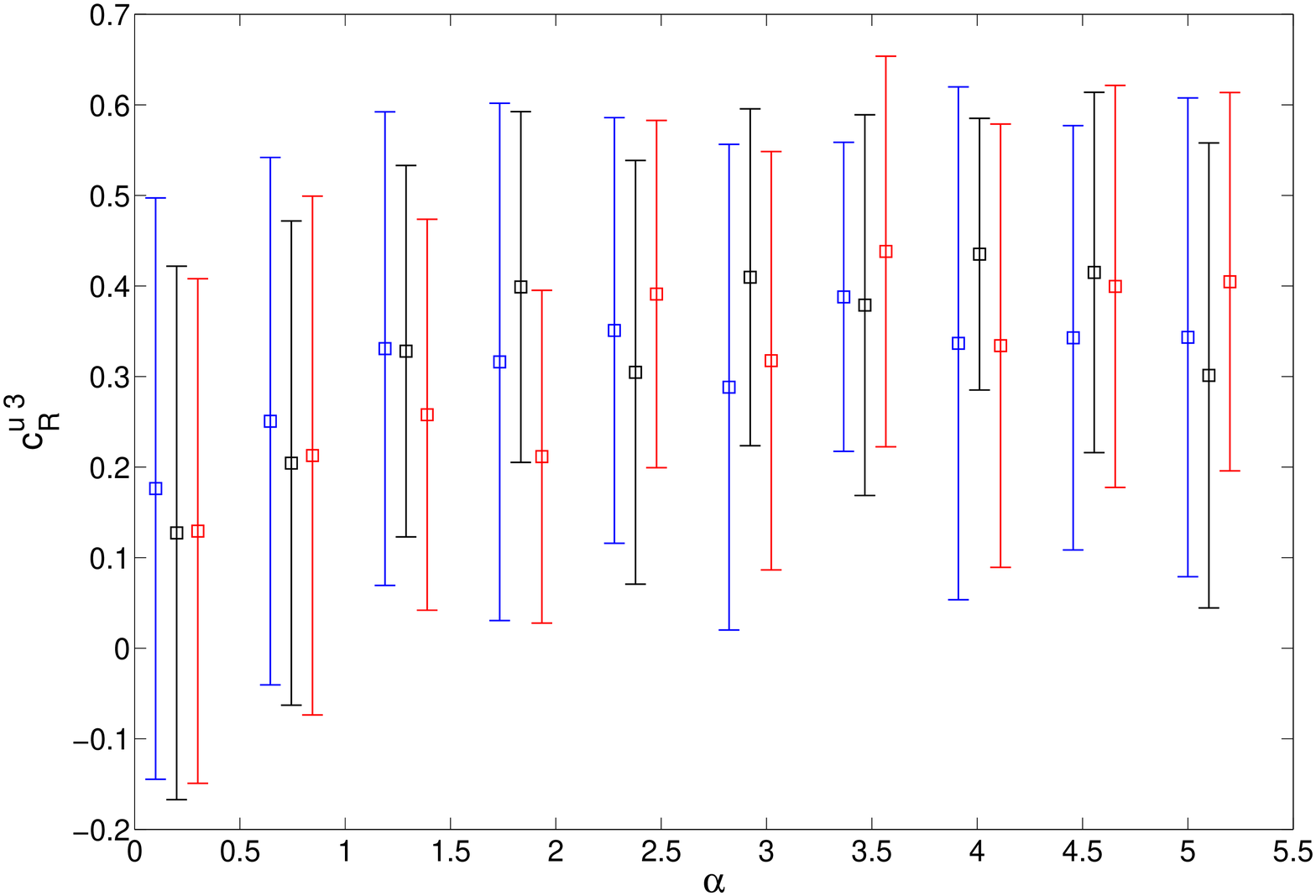}
        }
        \subfigure[$c_R^{u\, 2}$]{%
           \label{fig:cuR2}
           \includegraphics[width=0.45\textwidth]{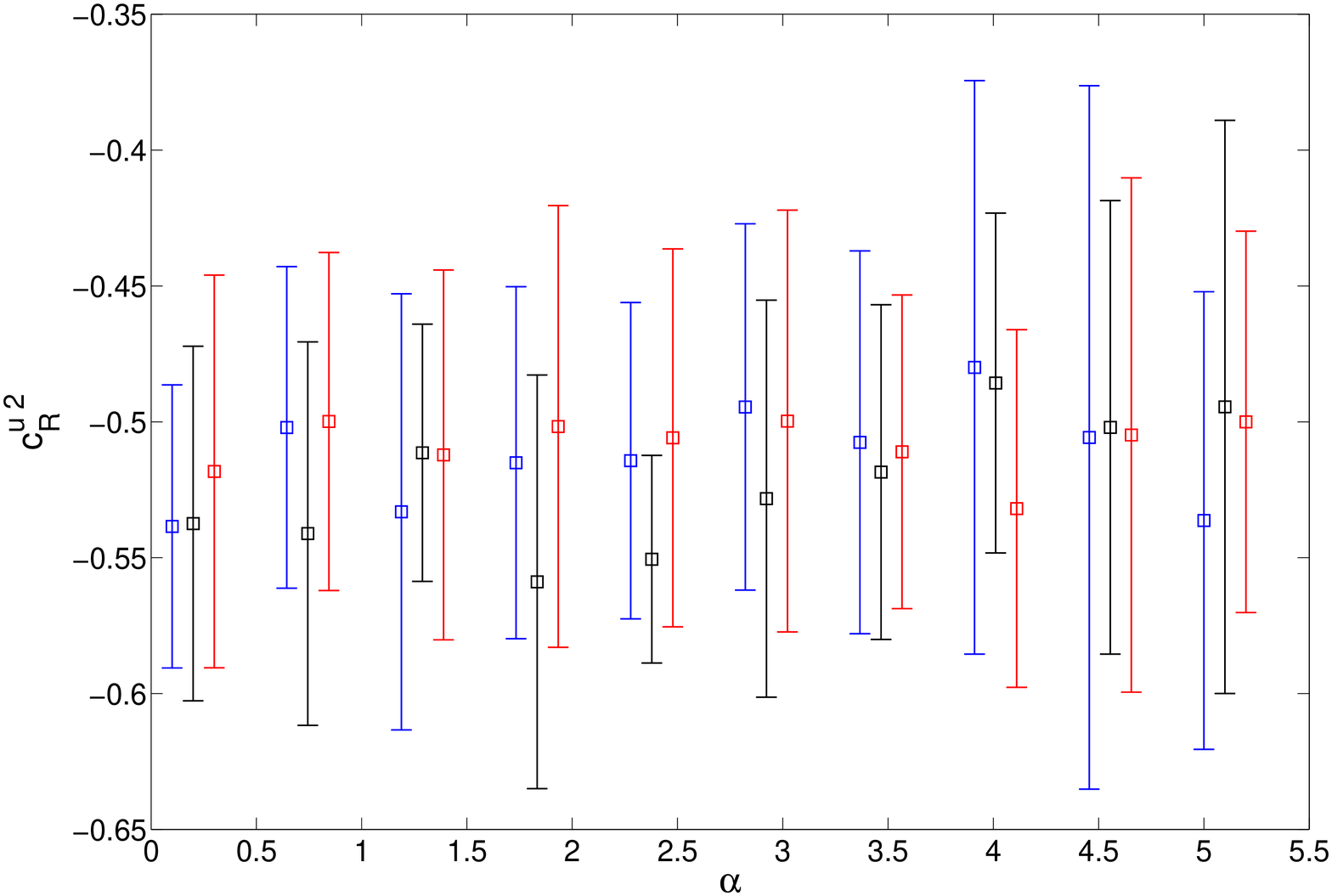}
        }\\
         \subfigure[$c_R^{d\, 3}$]{%
            \label{fig:cdR3}
            \includegraphics[width=0.45\textwidth]{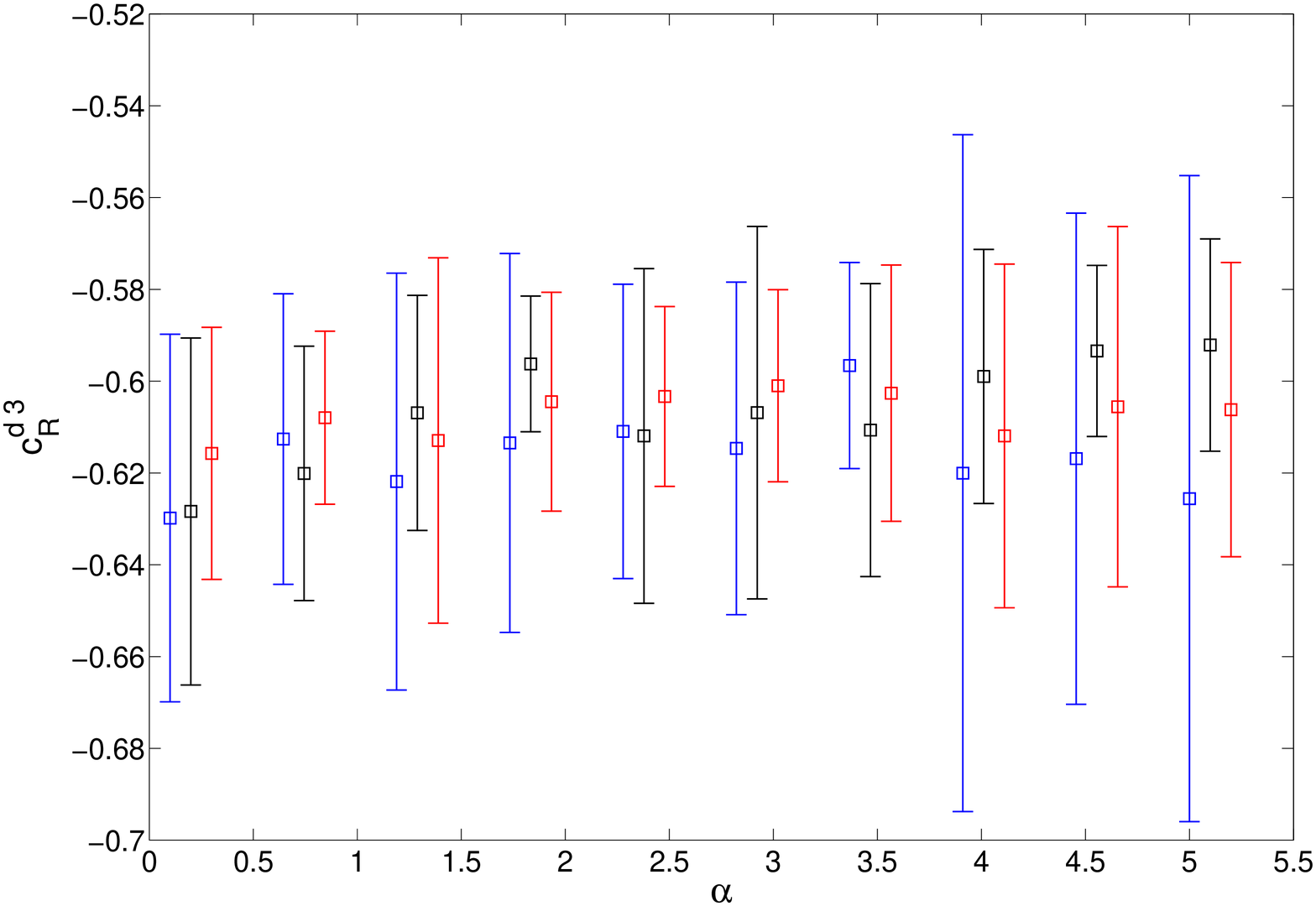}
        }
        \subfigure[$c_R^{d\, 2}$]{%
           \label{fig:cDR2}
           \includegraphics[width=0.45\textwidth]{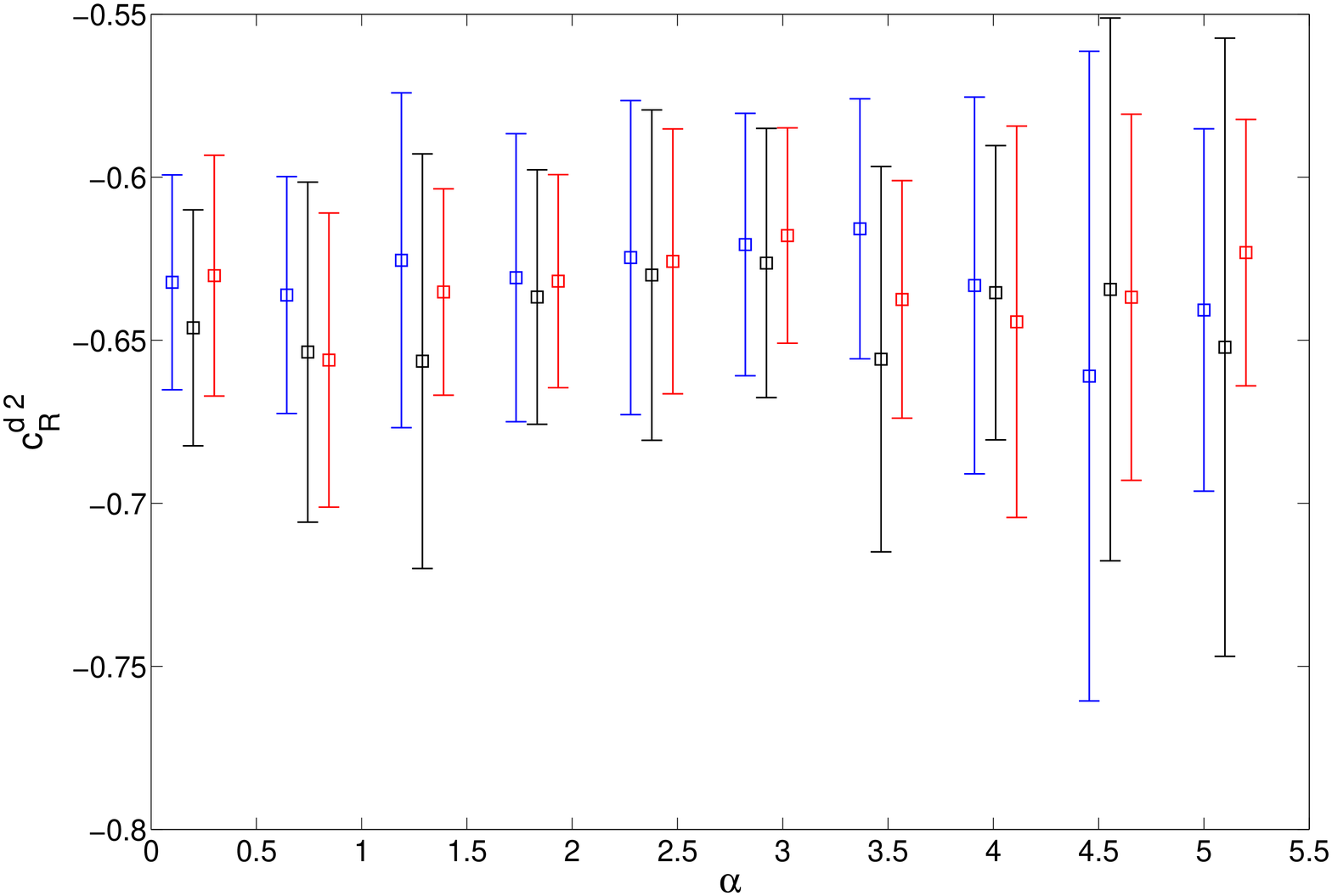}
        }\\
    \end{center}
    \caption{ The mean and standard deviations for the `preferred fermion' positions. When $c_L>0.5\,(<0.5)$ the left handed fermions profiles will be peaked towards the UV (IR) branes and likewise when $c_R<-0.5\,(>-0.5)$ the right handed fermions will be peaked towards the UV (IR). Here we have considered the RS model (black) and the modified metric with $k\Delta=1$ and $v=10$ (blue) and $v=3$ (red). The light quark positions do not demonstrate as significant an $\alpha$ dependence. $\Omega=10^{15}$ and $M_{\rm{KK}}=2$ TeV. } \label{fig:FermPso}
\end{figure}

However none of the above papers investigated the $\alpha$ dependence of this shift in the fermion position. A priori it is not obvious how the fermion positions are effected since it is based on two conflicting factors. Notably the `maximum fermion mass' decreases as $\alpha$ is reduced (see figure \ref{fig:MaxMass}) but the gradient of the slope in figure \ref{fig:BulkHiggs} also reduces. Hence this must be calculated explicitly. To do so we minimise a function that inputs the nine quark bulk mass parameters ($c_L$, $c^u_R$ and $c_R^d$) and computes the $\chi^2$ value for the mean quark masses, mean CKM mixing angles and mean Jarlskog invariant, taken over 5000 anarchic Yukawa matrices with $\frac{1}{3}\leqslant |\tilde{Y}|\leqslant 3$. The quark masses are run down to the mass of the first KK gluon from the $2$ GeV values \cite{Nakamura:2010zzi},
\begin{eqnarray}
m_u=2.4\pm 0.7\mbox{ MeV}\hspace{1cm}m_c=1.29^{+0.05}_{-0.11}\mbox{ GeV}\hspace{1cm} m_t=173\pm 1.5\mbox{ GeV}\nonumber\\
m_d=4.9\pm 0.8\mbox{ MeV}\hspace{1.0cm}m_s=100^{+30}_{-20}\mbox{ MeV}\hspace{1.3cm} m_b=4.2^{+0.18}_{-0.06}\mbox{ GeV}\hspace{0.2cm}\nonumber
\end{eqnarray}     
and mixing angles used are \cite{Nakamura:2010zzi, Bona:2007vi}
\begin{displaymath}
V_{us}=0.2254\pm0.00065\hspace{0.5cm}V_{cb}=0.0408\pm0.00045\hspace{0.5cm}V_{ub}=0.00376\pm0.0002\hspace{0.5cm}J=2.91^{+0.19}_{-0.11}\times 10^{-5}.
\end{displaymath}
Due to the size of the parameter space, such a minimisation routine will typically find a local minimum and not a global minimum. Hence this is repeated 200 times with random initial guesses in the ranges $c_L=[0.66\pm0.1,\; 0.58\pm 0.1,\; 0.4\pm0.2]$, $c_R^u=[-0.6\pm0.1,\; -0.51\pm 0.1,\; 0.2\pm0.5]$ and $c_R^d=[-0.66\pm0.1,\; -0.64\pm0.1,\; -0.59\pm0.1]$. Of these 200 configurations, the 40 best $\chi^2$ values are taken and the mean bulk mass parameters have been plotted in figure \ref{fig:FermPso} for different values of $\alpha$. Although some prior knowledge of warped extra dimensions has been used in choosing the initial guesses, it is still found that all the bulk mass parameters do still converge to a preferred value.  

Inevitably, due to the size of the parameter space, there are large numerical uncertainties associated with these preferred fermion positions. Also one could argue that a better analysis would include more observables such as $\epsilon_K$ and the Z partial decay width, $R_b$, in the fit. These would slightly shift the preferred values but it is suspected that the basic result would still hold\footnote{The $\alpha$ dependence of the fermion position is dominated by the gradient of the slope and the position of the `maximum fermion mass' in figure \ref{fig:BulkHiggs}. Including additional observables would not change this but would rather change the extent to which a split fermion scenario was favoured. For example the inclusion of $\epsilon_K$ would favour a shift in $c_L$ towards the IR and $c_R^{d}$ towards the UV \cite{Archer:2011bk}. While the inclusion of $R_b$ would disfavour a split fermion scenario.}. The result being that, despite these uncertainties, there is a generic trend for the fermions to sit further towards the UV for small values of $\alpha$. This would lead to their couplings being more universal and hence result in a reduction in the constraints from not just $\epsilon_K$ but a significant number of the $\Delta F=2$ and $\Delta F=1$ processes. One can see from the analysis made in \cite{Cabrer:2011qb, Agashe:2008uz, Archer:2011bk} that relatively small changes in the fermion positions can result in quite significant reductions in these constraints.  

Finally it should be pointed out that another source of flavour violation in these models arises from Higgs mediated FCNC's arising from the exchange of a KK fermion. This has not been considered here since it was studied in \cite{Azatov:2009na} and found to be largely independent of $\alpha$.

\section{Conclusions.}\label{sect:Conc}
The primary focus of this paper has been to investigate the phenomenological changes, in RS type scenarios, as one changes the exponent of a bulk Higgs VEV. The motivation for doing so is that, when the exponent is close to two and one assumes order one Yukawa couplings, the range of fermion zero mode masses is in remarkable agreement with the observed fermion mass hierarchy. After first introducing the model, it was demonstrated that, when brane localised potentials exist, then non tuned Higgs VEVs will not be flat, but will be peaked towards the IR. Hence bulk Higgs scenarios still offer a potential resolution to the gauge hierarchy problem. However, models with a bulk Higgs, gain an additional positive contribution to the $|\Phi|^2$ term in the effective potential which may result in EW symmetry not being broken for large values of the Higgs exponent, $\alpha$. Next the fermion mass hierarchy was considered and it was found that (assuming order one Yukawa couplings) the Dirac mass term of the fermion zero modes stretch from the EW scale to a factor of about $\Omega^{-1-\alpha}$ below it, see figure \ref{MinMaxMass} for the calculated values. We then proceeded to demonstrate that a considerable number of the phenomenological constraints will be at a minimum for smaller values of $\alpha$. In particular it is shown, for generic Higgs potentials and generic geometries, that the flatter the Higgs VEV the smaller the contribution to the $T$ parameter. It is also shown that a large class of constraints from flavour physics will be reduced for smaller values of $\alpha$. This is related to three effects. Firstly for the RS model the size of the pseudo-scalar coupling to the SM fermion is enhanced as one increases $\alpha$. This coupling is not flavour diagonal and would give rise to pseudo-scalar mediated FCNC's. This effect is significantly reduced in spaces with a modified metric. Secondly the non-universality of the $W$ and $Z$ coupling to SM fermions is reduced for smaller value of $\alpha$. Thirdly it is found that there is a general trend for the fermions to sit further towards the UV for smaller values of $\alpha$. This would result in their couplings to KK gauge fields being more universal with respect to flavour. 

All of the above motivations seem to favour a small value of $\alpha$ and are independent of the requirement of generating the correct fermion mass hierarchy. One also has a potential lower bound of $\alpha\geqslant 0$ coming from the Breitenlohner-Freedman bound, although this is not strictly applicable here. However the RS model is not a UV complete theory and it is plausible that attempts to embed such a model in a more fundamental theory could result in the Breitenlohner-Freedman bound being applicable. Another lower bound arises when one considers the 4D dual theory. The corresponding Higgs operator, $\mathcal{O}$, would have a scaling dimension of $2+\alpha$. In order to offer a potential resolution to the gauge hierarchy problem one requires that the operator $\mathcal{O}^\dag\mathcal{O}$ is not relevant which in turn implies that $\alpha\geqslant 0$ \cite{Luty:2004ye}. Hence it is compelling to saturate these two bounds, for the largely phenomenological reasons given in this paper and claim that the optimal value of $\alpha$ is zero (or equivalently the optimal exponent of the Higgs VEV and scaling dimension of the Higgs operator is two). 

Even if one accepts that $\alpha$ should be close to zero, then more work is still required before the fermion mass hierarchy can be understood. In particular, models with a bulk Higgs offer no explanation of, for example, why the leptons are lighter than the quarks, or why there is a six orders of magnitude gap between the electron mass and the neutrino masses. We have also focused on the Dirac mass term. It was argued in \cite{Agashe:2008fe}, that in RS type scenarios constraints, from lepton flavour violation, favour Dirac neutrinos over Majorana neutrinos. However there is no symmetry forbidding a Majorana mass term. The validity of this possible explanation of fermion mass hierarchy is conditional on such mass terms being either forbidden or existing at the same mass scale as the Dirac term. There are a number of possibilities for achieving this but all would require an extension of this minimal model and so we do not consider them here.     

Further work is also required in order to fully understand the phenomenological implications of both modifying the geometry and the pseudo-scalars. With regards to the modification of the geometry, although here we have not fully explored the possible parameter space, we still find significant phenomenological changes relative to the RS model. In particular it is found that, while the couplings do not change significantly, the relationship, between the curvature and size of the extra-dimension and the KK mass eigenvalues, does change. It is believed that this is partly responsible for the reduction in the EW constraints \cite{Cabrer:2011vu, Cabrer:2011fb, Carmona:2011ib}. With regards to the additional scalar degrees of freedom, assuming the LHC is able to reach the KK scale, then the prediction of a pseudo-scalar, for each $W^{\prime}$ and $Z^{\prime}$, makes models with a bulk Higgs very falsifiable. Although again we must leave a full study of such scalars to future work. Nevertheless, it is believed that there are a number of reasons for considering models with a bulk Higgs, an interesting extension of the description of flavour that already exists in the RS model.   

\section*{Acknowledgements.}
I am very grateful to Kristian McDonald for a number of useful discussions and comments. This work has also benefited from a number of useful discussions with Mathias Neubert, Susanne Westhoff and Thomas Flacke. This research was supported by the grant 05H09UME of the German Federal Ministry for Education and Research (BMBF).

\appendix

\section{Appendix}
In this appendix we shall derive many of the results used throughout this paper. Here we will work with a generic 5D warped space of the form,
\begin{equation}
\label{genericMetric }
ds^2=a(r)^2\eta^{\mu\nu}dx_\mu dx_\nu-b(r)^2dr^2,
\end{equation} 
with $r\in [r_{\rm{UV}}, r_{\rm{IR}}]$. Without loss of generality, $b(r)$ can always be set to one with the coordinate transformation $r\rightarrow\tilde{r}=\int^{r}_{c} b(\hat{r})d\hat{r}$. However, by not doing so, one can analytically express a greater range of spaces and also one can easily modify the following expressions in order to use a conformally flat metric.

\subsection{Electroweak Symmetry Breaking with a Bulk Higgs.}

Let us begin by considering a bulk $\rm{SU}(2)\times\rm{U}(1)$ gauge symmetry in addition to a bulk Higgs, $\Phi$,
\begin{equation}
\label{ }
S=\int d^5x\;\sqrt{G}\left (-\frac{1}{4}F_{MN}^a F_a^{MN}-\frac{1}{4}B_{MN}B^{MN}+|D_M\Phi|^2-V(\Phi)\right ).
\end{equation} 
Where the covariant derivative is given by $D_M=\partial_M-igA_M^a\tau^a-i\frac{1}{2}g^{\prime}B_M$, with the three $\rm{SU}(2)$ generators $\tau^a=\sigma^a/2$ and $a=\{1,2,3\}$. After spontaneous symmetry breaking the Higgs would acquire a non zero VEV, $\langle \Phi \rangle=\frac{1}{\sqrt{2}}\left(\begin{array}{c}0 \\h(r)\end{array}\right)$, where $h$ would be the solution of
\begin{equation}
\label{hVEVeom}
\partial_5\left (a^4b^{-1}\partial_5 h\right )-a^4b\partial_\Phi V(\Phi)|_{\Phi=h}=0.
\end{equation}
Clearly in order to actually break EW symmetry one must have a potential that does not admit $h=0$ as a solution. This can be done by either using the bulk potential, $V(\Phi)$, or by imposing non-trivial boundary conditions.  We can now expand $\Phi$ around $h$ such that,
\begin{equation}
\label{PHIDEF}
\Phi(x,r)=\frac{1}{\sqrt{2}}\left(\begin{array}{c}\pi_1(x,r)+i\pi_2(x,r) \\h(r)+H(x,r)+i\pi_3(x,r)\end{array}\right),
\end{equation}
and make the usual field redefinitions,
\begin{eqnarray}
W_M^{\pm}=\frac{1}{\sqrt{2}}\left (A_M^1\mp iA_M^{2}\right )\hspace{1cm}Z_M=\frac{1}{\sqrt{g^2+g^{\prime\,2}}}\left (gA_M^3-g^{\prime}B_M\right ) \hspace{1cm}
A_M=\frac{1}{\sqrt{g^2+g^{\prime\,2}}}\left (g^{\prime}A_M^3+gB_M\right ).
\end{eqnarray}
The Higgs particle itself will satisfy
\begin{equation}
\label{HiggsPartODE}
a^2b\partial_\mu\partial^\mu H-\partial_5(a^4b^{-1}\partial_5H)+a^4b\partial_\Phi V(\Phi)|_{\Phi=H}=0.
\end{equation}
It is also useful to define the quantities
\begin{equation}
\label{ MWZdef}
M_Z(r)\equiv\frac{\sqrt{g^2+g^{\prime\,2}}\,h(r)}{2}\hspace{0.5cm}\mbox{and}\hspace{0.5cm} M_W(r)\equiv \frac{g\,h(r)}{2}.
\end{equation}
These, of course, should not be confused with the 4D W and Z masses. We must now include a gauge fixing term which is chosen in order to cancel the mixing between the 4D gauge fields, $A_\mu^a$ and the 4D scalars, $A_5^a$ and $\pi_a$. In particular, working in the $R_\xi$ gauge, we introduce the gauge fixing term
\begin{eqnarray}
\mathcal{L}_{G.F.}=-\frac{b}{2\xi}\left (\partial_\mu Z^\mu-\xi b^{-1}\left (\partial_5(a^2b^{-1}Z_5)+a^2bM_Z\pi_3\right )\right )^2-\frac{b}{2\xi}\left (\partial_\mu A_1^\mu-\xi b^{-1}\left (\partial_5(a^2b^{-1}A^1_5)-a^2bM_W\pi_2\right )\right )^2 \nonumber\\
 -\frac{b}{2\xi}\left (\partial_\mu A_2^\mu-\xi b^{-1}\left (\partial_5(a^2b^{-1}A^2_5)+a^2bM_W\pi_1\right )\right )^2.
\end{eqnarray}  
If, for the moment, we just focus on the $Z$ field then the equations of motion are found to be
\begin{eqnarray}
-b\left (\partial^\nu Z_{\nu\mu}+\frac{1}{\xi}\partial_\mu(\partial_\nu Z^\nu ) \right )+\partial_5\left (a^2b^{-1}\partial_5Z_\mu\right )-a^2bM_Z^2Z_\mu=0,  \\
\partial_\mu\partial^\mu Z_5+a^2M_Z\partial_5\pi_3-a^2M_Z\left (\frac{\partial_5h}{h}\right )\pi_3+a^2M_Z^2Z_5-\xi\partial_5\left (b^{-1}\left (\partial_5(a^2b^{-1}Z_5)+a^2bM_Z\pi_3\right )\right )=0,\label{Z5EOM}\\
\partial_\mu\partial^\mu \pi_3-a^{-2}b^{-1}\partial_5\left (a^4b^{-1}\partial_5\pi_3\right )-a^{-2}b^{-1}\partial_5\left (a^4b^{-1}M_ZZ_5\right )-a^2b^{-2}M_Z\left (\frac{\partial_5h}{h}\right )Z_5+a^2\partial_{\Phi} V(\Phi)|_{\Phi=\pi_3}\hspace{1cm}\nonumber\\+\xi M_Zb^{-1}\left (\partial_5(a^2b^{-1}Z_5)+a^2bM_Z\pi_3\right )=0.\label{PI3EOM}
\end{eqnarray} 
To find the masses of the 4D Z bosons, one would expand $Z_\mu$ in terms of orthogonal mass eigenstates, i.e. make a KK decomposition, $Z_\mu(x,r)=\sum_nf_n^{(Z)}(r)Z_\mu^{(n)}(x)$ such that $\int dr\, bf_n^{(Z)}f_m^{(Z)}=\delta_{nm}$. The masses would then be defined by the 4D equations of motion, $\partial^\nu Z^{(n)}_{\nu\mu}+\frac{1}{\xi}\partial_\mu(\partial^\nu Z^{(n)}_\nu )=m_n^{(Z)\,2}Z_\mu^{(n)}$, and found by solving for the profiles, 
\begin{equation}
\label{ZEOM}
\partial_5\left (a^2b^{-1}\partial_5f_n^{(Z)}\right )-a^2bM_Z^2f_n^{(Z)}+bm_n^{(Z)\,2}f_n^{(Z)}=0.
\end{equation}
The $W$ and photon profiles, $f_n^{(W)}$ and $f_n^{(\gamma)}$ are found using the same equation but with $M_Z$ replaced by $M_W$ and $0$ respectively.

\subsection{The Gauge-Goldstone Boson Equivalence Theorem Cross Check.}
There are now two remaining scalar degrees of freedom. One will be eaten by the longitudinal polarisation states of the 4D massive $Z$ fields and will be the unphysical Goldstone boson. The other will be a physical pseudo-scalar.  To find the Goldstone bosons we simply note that such fields must have a completely gauge dependent mass. Such that in the unitary gauge ($\xi\rightarrow \infty$) the Goldstone bosons become infinitely heavy and hence should be considered unphysical. From (\ref{Z5EOM}) and (\ref{PI3EOM}) it is straightforward to see that the Goldstone boson should be given by
\begin{equation}
\label{GoldZDef}
\mathcal{G}_Z=b^{-1}\left (\partial_5\left (a^2b^{-1}Z_5\right )+a^2bM_Z\pi_3\right ).
\end{equation}   
However in order for the gauge-Goldstone boson equivalence theorem to hold, in the 4D theory, one necessarily requires that the profiles of the Goldstone boson, $f_n^{(\mathcal{G}_Z)}$, should match those of the Z boson, $f_n^{(Z)}$. Otherwise the 4D effective couplings would be different for the two fields and hence the scattering amplitudes would not be the same. This equivalence has been found in other extra dimensional scenarios, such as when a gauge field propagates in more than five dimensions \cite{McDonald:2009hf}. To check that the profiles do match we must find the equations of motion for the Goldstone bosons. In particular, taking $\partial_5(a^2b^{-1}( \ref{Z5EOM}))$, adding $a^2bM_Z(\ref{PI3EOM})$ and noting that $\partial_5(a^4b^{-1}M_Z^2Z_5)-M_Z\partial_5(a^4b^{-1}M_ZZ_5)-a^4b^{-1}M_Z^2(\frac{\partial_5h}{h})Z_5=0$ gives
\begin{displaymath}
\partial_\mu\partial^\mu\mathcal{G}_Z-\frac{\sqrt{g^2+g^{\prime\;2}}}{2}\partial_5\left (a^4b^{-1}\partial_5h\right )\pi_3+a^4bM_Z\partial_{\Phi} V(\Phi)|_{\Phi=\pi_3}-\xi b^{-1}\partial_5\left (a^2b^{-1}\partial_5 \mathcal{G}_Z\right )-\xi a^2M_Z^2\mathcal{G}_Z=0.
\end{displaymath}
For all Higgs potentials, considered in this paper, the second and third gauge independent terms will cancel using (\ref{hVEVeom}). One would anticipate that such a cancellation would occur for all Lorentz and gauge invariant potentials, but this is not verified here. With such a cancellation, one can proceed to make a KK decomposition $\mathcal{G}_Z=\sum_n f_n^{(\mathcal{G}_Z)}(r)\mathcal{G}^{(n)}_Z(x)$ such that $\partial_\mu\partial^\mu\mathcal{G}^{(n)}_Z=-\xi m_n^{(\mathcal{G}_Z)\,2}\mathcal{G}_Z^{(n)}$  and the profiles will be given by
\begin{equation}
\label{ }
\partial_5\left (a^2b^{-1}\partial_5f_n^{(\mathcal{G}_Z)}\right )-a^2bM_Z^2f_n^{(\mathcal{G}_Z)}+bm_n^{(\mathcal{G}_Z)\,2}f_n^{(\mathcal{G}_Z)}=0,
\end{equation}
in agreement with (\ref{ZEOM}).

\subsection{The Physical Pseudo-Scalars.}
To find the remaining physical degree of freedom we must take a combination of (\ref{Z5EOM}) and (\ref{PI3EOM}) such that the mass term is gauge independent. This can be done by adding (\ref{Z5EOM}) to $\partial_5(M_Z^{-1}(\ref{PI3EOM}))$ which, after a little algebra, gives
\begin{equation}
\label{pseudoEOMZ}
\partial_\mu\partial^\mu \phi_Z-\partial_5\left (a^{-2}b^{-1}M_Z^{-2}\partial_5\left (a^4b^{-1}M_Z^2\phi_Z\right )\right )+a^2M_Z^2\phi_Z=0.
\end{equation}
We have again used the cancellation $h^{-1}\partial_5(a^4b^{-1}\partial_5h)+a^4b\partial_{\Phi}V(\Phi)|_{\Phi=\pi_3}=0$, as well as defining
\begin{equation}
\label{ phiZDef}
\phi_Z\equiv \partial_5\left (M_Z^{-1}\pi_3\right )+Z_5.
\end{equation}
This equation agrees with equivalent expressions derived previously in \cite{Cabrer:2011fb, Falkowski:2008fz}. In order to find the effective  action for these pseudo-scalars, we again make a KK decomposition,  $\phi_Z(x,r)=\sum_n f_n^{(\phi_Z)}(r)\phi_Z^{(n)}(x)$, such that $\partial_\mu\partial^\mu\phi_Z^{(n)}=-m_n^{(\phi_Z)\;2}\phi_Z^{(n)}$. This allows us to invert (\ref{GoldZDef}) and (\ref{ phiZDef}) to get
\begin{eqnarray}
\pi_3=\sum_n \left (-\frac{M_Z^{-1}a^{-2}b^{-1}\partial_5\left (a^4b^{-1}M_Z^2f_n^{(\phi_Z)}\right )}{m_n^{(\phi_Z)\;2}}\phi_Z^{(n)}+\frac{M_Zf_n^{(\mathcal{G}_Z)}}{m_n^{(\mathcal{G}_Z)\,2}}\mathcal{G}^{(n)}_Z\right ) \label{pi3mix}\\
Z_5=\sum_n\left (\frac{a^2M_Z^2f_n^{(\phi_Z)}}{m_n^{(\phi_Z)\;2}}\phi_Z^{(n)}-\frac{\partial_5f_n^{(\mathcal{G}_Z)}}{m_n^{(\mathcal{G}_Z)\,2}}\mathcal{G}^{(n)}_Z\right ).\label{Z5mix}
\end{eqnarray}
For the bulk of this paper we shall work in the unitary gauge, in which the Goldstone bosons are infinitely heavy and can be neglected. The final step is to ensure the mass eigenstates are canonically normalised which is achieved with the orthogonality relation
\begin{equation}
\label{PhiOrthogRel}
\int dr\; a^4b^{-1}\frac{M_Z^2}{m_n^{(\phi_Z)\;2}} f_n^{(\phi_Z)}f_m^{(\phi_Z)}=\delta_{nm}.
\end{equation}

\subsection{The W Boson.}
It is straight forward to repeat this analysis for the W field, where it is found that the two Goldstone bosons are given by
\begin{equation}
\label{ }
\mathcal{G}_W^{1}=b^{-1}\partial_5\left (a^2b^{-1}A_5^1\right )-a^2bM_W\pi_2\hspace{0.8cm}\mbox{and}\hspace{0.8cm}\mathcal{G}_W^{2}=b^{-1}\partial_5\left (a^2b^{-1}A_5^2\right )+a^2bM_W\pi_1.
\end{equation}
While the two physical pseudo-scalars are found to be
\begin{equation}
\label{PhiWdef}
\phi_W^{1}=\partial_5(M_W^{-1}\pi_2)-A_5^1\hspace{0.8cm}\mbox{and}\hspace{0.8cm}\phi_W^{2}=\partial_5(M_W^{-1}\pi_1)+A_5^2.
\end{equation}
In practice there will be a further mixing such that the charged pseudo-scalars are given by combinations of $\pi^{\pm}=\frac{1}{\sqrt{2}}(\pi_1\pm i\pi_2)$ and $W_5^{\pm}=\frac{1}{\sqrt{2}}(A_5^1\mp iA_5^2)$. After making analogous KK decompositions, as those used for the Z field, the equations of motion are analogously
\begin{eqnarray}
\partial_\mu\partial^\mu\phi^{1,2}_W-\partial_5(a^{-2}b^{-1}M_W^{-2}\partial_5(a^4b^{-1}M_W^2\phi_W^{1,2}))+a^2M_W^2\phi_W^{1,2}=0\label{pseudoEOMW}\\
\partial_\mu\partial^\mu\mathcal{G}_W^{1,2}-\xi b^{-1}\partial_5\left (a^2b^{-1}\partial_5 \mathcal{G}^{1,2}_W\right )+\xi a^2M_W^2\mathcal{G}_W^{1,2}=0.
\end{eqnarray} 
The effective action can be found by substituting for
\begin{eqnarray}
\pi_{1,2}=\sum_n \left (-\frac{M_W^{-1}a^{-2}b^{-1}\partial_5\left (a^4b^{-1}M_W^2f_n^{(\phi_W)}\right )}{m_n^{(\phi_W)\;2}}\phi_W^{2,1\,(n)}\pm\frac{M_Wf_n^{(\mathcal{G}_W)}}{m_n^{(\mathcal{G}_W)\,2}}\mathcal{G}^{2,1\;(n)}_W\right )\label{piWMix} \\
A^{1,2}_5=\sum_n\left (\mp\frac{a^2M_W^2f_n^{(\phi_W)}}{m_n^{(\phi_W)\;2}}\phi_W^{1,2\,(n)}-\frac{\partial_5f_n^{(\mathcal{G}_W)}}{m_n^{(\mathcal{G}_W)\,2}}\mathcal{G}^{1,2\,(n)}_W\right ).\label{W5mix}
\end{eqnarray}

\subsection{The Fermion Profile.}
For completeness, we shall briefly derive the widely known expressions for the fermion profiles. Beginning with the action for a massive Dirac spinor, $\Psi$, in 5D
\begin{equation}
\label{FermAction}
S=\int d^5x \sqrt{G}\left (i\bar{\Psi}\Gamma^M\nabla_M\Psi-M_\Psi\bar{\Psi}\Psi\right )
\end{equation}
where the Dirac Matrices in curved space $\Gamma^M=E_A^M\gamma^A$ are related to Dirac matrices in flat space, $\gamma^A$, by the f\"{u}nfbein $E_A^M=\mathrm{diag}(a^{-1},a^{-1},a^{-1},a^{-1},b^{-1})$. While the covariant derivative, $\nabla_M=D_M+\omega_M$, includes the spin connection $\omega_M=(\frac{1}{2}b^{-1}\partial_5a\,\gamma_5\gamma_\mu,0)$. Splitting $\Psi=\psi_L+\psi_R$ such that $i\gamma_5\psi_{L,R}=\mp \psi_{L,R}$ and making the KK decomposition $\psi_{L,R}=\sum_na^{-2}f_n^{(\psi_{L,R})}(r)\psi_{L,R}^{(n)}(x)$ such that $(i\gamma^\mu\partial_\mu-m_n^{(\psi)})\Psi^{(n)}=0$. This then yields the coupled equations of motion
\begin{equation}
\label{FermEOM}
\partial_5 f_n^{(\psi_R)}+bM_\Psi f_n^{(\psi_R)}=\frac{b}{a}m_n^{(\psi)} f_n^{(\psi_L)}\hspace{0.8cm}\mbox{and}\hspace{0.8cm}-\partial_5f_n^{(\psi_L)}+bM_\Psi f_n^{(\psi_L)}=\frac{b}{a}m_n^{(\psi)} f_n^{(\psi_R)}.
\end{equation}
It is then straight forward to solve for the fermion zero mode profile
\begin{equation}
\label{fermProf}
f_0^{(\psi_{L,R})}(r)=\frac{\exp\left (\pm\int_c^r d\tilde{r}\,b(\tilde{r})M_\Psi\right )}{\sqrt{\int dr\,\frac{b}{a}\exp\left (\pm2\int_c^r d\tilde{r}\,b(\tilde{r})M_\Psi\right )}}.
\end{equation} 
It is common to parameterise the bulk mass term such that $M_\Psi=-ck$ where $c\sim \mathcal{O}(1)$.

\bibliographystyle{JHEP}

\bibliography{bibliography}   

\end{document}